\documentclass[prd,preprint,aps,showkeys,tightenlines,floatfix,showpacs,preprintnumbers,nofootinbib,eqsecnum]{revtex4-1}

\usepackage{silence}
\usepackage[dvips,final]{graphicx}
\usepackage{amssymb}
\usepackage{amsmath}
\usepackage{amsfonts}
\usepackage{pifont}
\usepackage{epsfig}
\usepackage{bm}
\usepackage{comment}
\usepackage{float}
\usepackage[justification=Justified]{caption}
\usepackage{subcaption}
\usepackage{hhline}
\usepackage[dvipsnames,table,xcdraw]{xcolor}
\usepackage{nth}
\usepackage{cancel}
\usepackage{fontawesome5} 

\usepackage{multirow}
\usepackage{bbold}
\usepackage{hyperref}
\hypersetup{
    pdftitle={Gravitational waves from first order phase transitions in a leptoquark model},
    pdfsubject={hep-ph, hep-th, gr-qc},
    pdfauthor={Marco~Finetti, Vasileios~Vatellis},
    pdfkeywords={BSM, leptoquark, first order phase transitions, gravitational waves},
    pdfnewwindow=true,    
    colorlinks=true,      
    linkcolor=blue,       
    citecolor=blue,       
    urlcolor=blue         
}
\usepackage{enumerate}
\usepackage{mathtools}
\usepackage{booktabs}
\usepackage{outlines} 

\usepackage{physics} 
\usepackage{makecell} 
\usepackage{booktabs}

\usepackage{cancel}

\newcommand{\U}[1]{\mathrm{U}(1)_{\mathrm{#1}}}			
\newcommand{\SU}[2]{\mathrm{SU}(#1)_{\mathrm{#2}}}		

\usepackage[capitalise]{cleveref}

\crefname{section}{Sec.}{Secs.}
\crefname{table}{Tab.}{Tabs.}
\crefname{figure}{Fig.}{Figs.}
\crefname{equation}{Eq.}{Eqs.}
\crefname{appendix}{Appendix\ }{Appendix\ }

\definecolor{bostonuniversityred}{rgb}{0.8, 0.0, 0.0}

\newcommand{\green}[0]{\color{green!50!black}}
\newcommand{\blue}[0]{\color{blue!99!black}}

\newcommand{\red}[0]{\color{red!90!black}}

\newcommand\varpm{\mathbin{\vcenter{\hbox{%
  \oalign{\hfil$\scriptstyle\hspace{-0.1ex}+\hspace{-0.1ex}$\hfil\cr
          \noalign{\kern-.5ex}
          $\scriptscriptstyle({-})$\cr}%
}}}}

\DeclareSymbolFont{myletters}{OML}{ztmcm}{m}{it}
\DeclareMathSymbol{\uplambda}{\mathord}{myletters}{"15}
\newcommand{\logbar}[1]{\overline{\log}\, #1}
\newcommand{\ms}{m_{S^{2/3}}}
\newcommand{\mS}[1]{m_{S_{#1}^{1/3}}}
\newcommand{\Sa}[1]{{S_{#1}^{1/3}}}

\begin{document}

\title{Gravitational waves from color restoration in a leptoquark model of radiative neutrino masses}

\author{Mårten Bertenstam$^{2}$}
\email{marten.bertenstam@fysik.lu.se}

\author{Marco Finetti$^{1}$}
\email{mfinetti@ua.pt}

\author{António~P.~Morais$^{1}$}
\email{aapmorais@ua.pt}

\author{Roman Pasechnik$^{2}$}
\email{roman.pasechnik@fysik.lu.se}

\author{Johan Rathsman$^{2}$}
\email{johan.rathsman@fysik.lu.se}

\affiliation{
\\
{$^1$\sl 
Departamento de F\'{i}sica da Universidade de Aveiro and Centre  for  Research  and  Development in Mathematics and Applications (CIDMA), Campus de Santiago 3810-183 Aveiro, Portugal.
}\\
{$^2$\sl
Department of Physics, Lund University, 221 00 Lund, Sweden.
}\\
}

\begin{abstract}
We study the first-order phase transitions and the emerging stochastic gravitational wave spectrum in a minimal leptoquark extension of the Standard Model that explains active neutrino oscillation data while satisfying current flavor physics constraints. This model exhibits diverse phase transition patterns, including color symmetry-breaking scenarios in the early Universe. Strong correlations between model parameters and gravitational-wave signals yield testable predictions for future experiments such as LISA, BBO, and DECIGO. Specifically, a detectable signal in the mHz–0.1~Hz frequency range features color-restoration and leptoquark masses near 1.5~TeV. With this article, we also present the first application in the literature of \texttt{Dratopi}. This is a soon-to-be-released tool for phase transition analysis using the dimensional reduction formalism, that interfaces the \texttt{DRalgo} package with \texttt{Python} and a slightly modified version of \texttt{CosmoTransitions}.
\end{abstract}

\maketitle
\tableofcontents

\newpage
\section{Introduction}

One of the most exciting prospects in the emerging field of gravitational-wave (GW) astrophysics \cite{LIGOScientific:2016aoc} is the possibility to observe a stochastic GW background (SGWB) originating from cosmological sources, such as first-order phase transitions (FOPTs) in the early Universe \cite{Kamionkowski:1993fg,Caprini:2015zlo,Caprini:2018mtu,Caprini_2020,Caprini:2024hue}. Although the Standard Model (SM) features a cross-over electroweak (EW) transition \cite{Kajantie:1995kf,Kajantie:1996mn,Gurtler:1997hr}, FOPTs are predicted to occur in various beyond the SM (BSM) scenarios \cite{Fromme:2006cm,Huang:2016cjm,Gould:2019qek}. Such transitions are typically motivated by the \emph{EW baryogenesis} mechanism \cite{Anderson:1991zb,Cohen:1993nk,Morrissey:2012db}, where a first-order EW phase transition (EWPT) pushes the plasma out of equilibrium creating the conditions necessary for the generation of the baryon asymmetry observed today. This process involves the spontaneous breaking of the EW symmetry, where the Higgs field acquires a non-zero vacuum expectation value (VEV).

Based on \cite{Freitas:2022gqs}, we investigate the possibility of breaking the  $\SU{3}{C}$  (\emph{color}) symmetry at finite temperatures in an extension of the SM involving two scalar leptoquarks (LQs), and explore the potential to detect the SGWB emergent from its restoration as the Universe cools down. We present, for the first time, a comprehensive analysis of a color-restoration cosmological FOPT within a framework that offers compelling insights from a particle physics perspective. Notably, \cite{Freitas:2022gqs} demonstrated how the considered LQ model --- which represents the minimal extension capable of radiatively generating Majorana neutrino masses \cite{Dorsner:2017wwn,AristizabalSierra:2007nf,Zhang:2021dgl,Pas:2015hca,Cai:2017jrq,Babu:2010vp,Cata:2019wbu} --- is consistent with $\mathcal{O}(100)$ flavor physics observables. In this study, we compare GW spectra --- obtained from the numerical scanning of the model's parameter space --- with the sensitivity of planned GW detectors: the Laser Interferometry Space Antenna (LISA) \cite{LISA:2017pwj,Colpi:2024xhw}, the DECi-hertz Interferometer Gravitational wave Observatory (DECIGO) \cite{Kawamura:2006up}, and the Big Bang Observer (BBO) \cite{Harry:2006fi}.

The hypothesis of color-breaking in the early Universe has been previously investigated in \cite{Patel:2013zla,Ramsey-Musolf:2017tgh,Chao:2021xqv}. Notably, a scenario similar to the one examined in this article was studied in \cite{Patel:2013zla}, which is focused on the mass-parameter space that accommodates color-breaking. Meanwhile, \cite{Ramsey-Musolf:2017tgh} explored multi-step phase transitions --- involving color-breaking --- that result in a purely EWPT, with implications for EW baryogenesis. Similarly, \cite{Chao:2021xqv} analyzed analogous scenarios in the context of EW baryogenesis with a focus on multi-step transitions where color-symmetry is broken and restored during the spontaneous breaking of the EW symmetry. In contrast, \cite{Fu:2022eun} considered cosmological phase transitions from a single, $\SU{2}{L}$-multiplet LQ model, focusing on the EWPT (where only the Higgs field acquires a VEV) and derived the primordial GW spectrum for a number of benchmark points.

The article is structured as follows. In \cref{sec:model} we present the LQ model in detail and discuss how to perform its matching to the SM. Then, \cref{sec:PT&GW} briefly introduces \emph{dimensional reduction} as a method to derive an effective 3d theory from the LQ model. We proceed by discussing general features of a cosmological phase transition and the SGWB it produces, motivating the specific choices made in this work. Furthermore, \cref{sec:numerics} describes the numerical results and routines implemented in this work. In particular, we present the SGWB peak amplitudes and frequencies and the corresponding thermodynamic parameters, discuss high-temperature perturbativity in the dimensional reduction paradigm, and categorize the various vacuum configurations obtained in the numerical analysis. Finally, \cref{sec:conclusions} summarizes the results and presents the conclusions of this work.

\section{Scalar leptoquark model} \label{sec:model}

Both neutrino masses and their mixing structure can be radiatively generated at the one-loop level by means of scalar LQs in the loops. This framework necessitates the presence of at least one LQ pair \cite{Chua:1999si,Mahanta:1999xd,Dorsner:2017wwn}, in addition to the SM-like Higgs doublet. Based on \cite{Freitas:2022gqs}, we consider a minimal LQ model of this type, featuring a pair of scalar LQs, commonly denoted as $\Tilde{R}_2$ and $S_1$, which correspond to a colored $\SU{2}{L}$ doublet and a colored $\SU{2}{L}$ singlet, respectively. To simplify the notation, we drop the indices, labeling the $\SU{2}{L}$ doublet and singlet as $R$ and $S$, respectively. In \cref{tab:charges} we present their transformation properties under the SM gauge group.
\begin{table}[ht]
    \centering
    \begin{tabular}{|c||c|c|c|}
    \hline
      Field & $\SU{3}{C}$ & $\SU{2}{L}$ & $\U{Y}$ \\ [0.5ex] 
     \hline\hline
        $H$ & $\mathbf{1}$ & $\mathbf{2}$ & 1/2 \\
     \hline
        $R$ & $\mathbf{3}$ & $\mathbf{2}$ & 1/6 \\
     \hline
        $S$ & $\mathbf{\overline{3}}$ & $\mathbf{1}$ & 1/3 \\
    \hline
    \end{tabular}
    \caption{\footnotesize Scalar field charges under the SM gauge group in the minimal LQ model.}
    \label{tab:charges}
\end{table}

\noindent The tree-level scalar potential reads
\begin{align}\label{eq:Vtree}
    \begin{split}
    V_{\text{tree}} =&\ \mu^2_H H^\dagger H  + \mu^2_R R^\dagger R  + \mu^2_S S^\dagger S \\
    & + \lambda_H (H^\dagger H)^2 + \lambda_R (R^\dagger R)^2 + \lambda_S (S^\dagger S)^2 \\
    & + g_{HS} (H^\dagger H)(S^\dagger S) + g_{HR} (H^\dagger H)(R^\dagger R) + g'_{HR} (H^\dagger R)(R^\dagger H) + g_{RS} (R^\dagger R )(S^\dagger S)  \\
    & + (a_1 R S H^\dagger + \text{h.c.})
    \end{split}    
\end{align}
Throughout this study, all model parameters are assumed to be real. This potential is symmetric under the simultaneous change of sign of any two fields ($e.g.$, $H\rightarrow-H$ and $R\rightarrow-R$) which allows us to impose two vacuum directions to be positive without loss of generality.

In addition to providing a mechanism for generating both Majorana masses and the mixing structure of neutrinos, it was shown in \cite{Freitas:2022gqs} that the model is consistent with flavor physics, complying with $\mathcal{O}(100)$ observables. The authors further emphasize that the same LQ model can offer explanations for several flavor anomalies, including those in $B$-physics, the muon anomalous magnetic moment and $W$-mass anomaly. However, recent findings suggest that these anomalies, particularly the last two, are likely not realized in nature. Various lattice results \cite{Borsanyi:2020mff,Ce:2022kxy} strongly suggest that the muon magnetic moment is consistent with the SM prediction. Similarly, the most recent $W$-mass measurement by the CMS collaboration \cite{ATLAS:2024erm,CMS:2024nau} also aligns well with the SM.

\subsection{Leptoquark mass spectrum} \label{sec:LQmasses}

\begin{figure}[ht]
    \centering
    \includegraphics[width=0.6\textwidth]{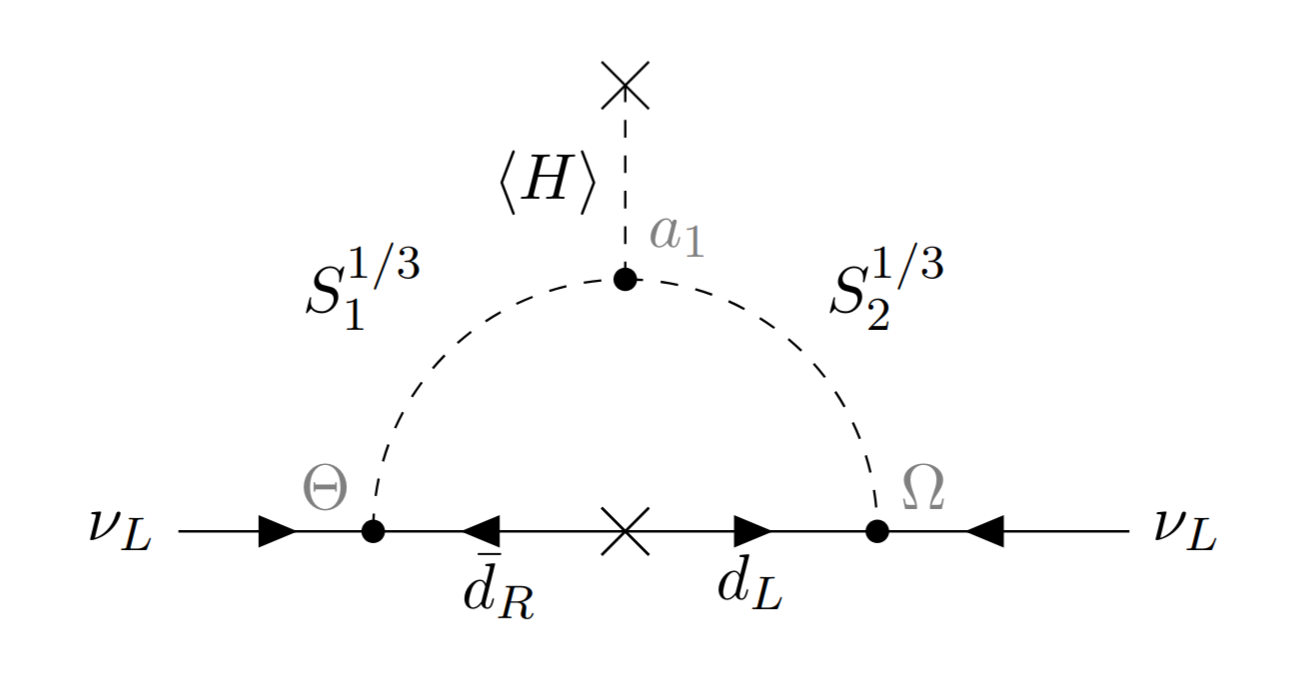}
    \caption{Majorana neutrino mass induced at the one-loop level by Higgs-leptoquark mixing.}
    \label{fig:neutrinoMass}
\end{figure}

Alongside the Higgs boson mass $m_H = \sqrt{\lambda_H}v_H \approx 125.10~\mathrm{GeV}$, the model introduces three additional masses, which at tree level are expressed as follows:
\begin{equation} 
\begin{aligned}\label{eq:lq_masses}
    \ms^2 &= \mu^2_R +
        \frac{1}{2} v_H^2 g_{HR} \,, \\
    \mS1^2 &= \frac{1}{2}\left(m_a^2 - m_b^2\right) \,, \\
    \mS2^2 &= \frac{1}{2}\left(m_a^2 + m_b^2\right) \,,
\end{aligned}
\end{equation}
where we have defined
\begin{equation}
\begin{aligned}
    m_a^2 &\equiv \mu^2_R + \mu^2_S + \frac{1}{2} \left(\Tilde{g}_{HR} + g_{HS}\right) v_H^2 \,, \\
        m_b^4 &\equiv \left[\mu_R^2-\mu_S^2+\frac{1}{2}(\Tilde{g}_{HR}-g_{HS})v_H^2 \right]^2 + 2a_1^2 v_H^2 \,,
\end{aligned}
\end{equation}
and $\Tilde{g}_{HR} \equiv g_{HR} + g_{HR}'$ for readability. We adopt the same notation for the mass eigenstates as in \cite{Freitas:2022gqs}, with the superscript indicating the respective electric charge. \cref{fig:neutrinoMass} illustrates the generation of Majorana neutrino masses through the Higgs-LQ mixing at the one-loop level, involving the $S_{1,2}^{1/3}$ mass eigenstates. Note that in the limit  $a_1 \to 0$, the neutrino masses vanish, highlighting the crucial role of the trilinear coupling.

The eigenstates $\Sa{1,2}$ arise from the mixing between one component of the $R$ doublet and the $S$ singlet, which depends on a non-zero $a_1$ coupling. This defines the mixing angle as\footnote{Up to a minus sign, this matches the definition in \cite{Freitas:2022gqs}.}
\begin{equation}\label{eq:mixingAngle}
    \sin{(2\theta)} = \frac{\sqrt{2}v_H a_1}{m_{S_2^{1/3}}-m_{S_1^{1/3}}}\ .
\end{equation}
Since all scalar sector parameters are real, we have $m_{S_2^{1/3}} \ge m_{S_1^{1/3}}$. Consequently, one can restrict the trilinear parameter $a_1$ to positive values, as negative sign can be removed by a redefinition of a field in the vertex leaving \cref{eq:Vtree} invariant. This enables us to choose $\theta\in [0,\pi/2]$ via \cref{eq:mixingAngle} without loss of generality.

\subsection{Matching to the Standard Model}
 
To ensure consistency with the SM at low energies, we match \cref{eq:Vtree} with the Higgs potential
\begin{equation}
    V_{\text{SM}}^{(0)} = \mu^2 (H^\dagger H) + \lambda (H^\dagger H)^2\ .
\end{equation}
To distinguish the Higgs sector parameters of the LQ model from those in the SM effective theory, we label the former with an $H$ subscript ($e.g.$, $\mu_H$, $\lambda_H$, $v_H$).
In the zero-momentum approximation it is sufficient to match the second and fourth derivatives of the SM potential to those of the one-loop LQ potential:
\begin{align}
    \frac{\partial^2 V_{\text{SM}}^{(0)}}{\partial H \partial H^\dagger} &= \frac{\partial^2 V_{\text{LQ}}^{(0+1)}}{\partial H \partial H^\dagger}\\[5pt]
    \frac{\partial^4 V_{\text{SM}}^{(0)}}{(\partial H)^2 (\partial H^\dagger)^2} &= \frac{\partial^4 V_{\text{LQ}}^{(0+1)}}{(\partial H)^2 (\partial H^\dagger)^2} \,.
\end{align}
The latter were computed using the generic expressions for derivatives of the effective one-loop potential developed in \cite{Camargo-Molina:2016moz}\footnote{The \texttt{Mathematica} notebook computing the one-loop derivatives can be found in the git repository \href{https://gitlab.com/mfinetti/dr_leptoquarks/-/blob/main/Mathematica\%20notebooks/LQ_match.nb}{\faGitlab}.}.

The resulting matching conditions for the SM parameters $\mu$ and $\lambda$ are 
\begin{flalign}\label{eq:matching_mu}
    \nonumber
    \mu^2 = \mu^2_H + \frac{3}{32\pi^2}\biggl[
        & \cancel{6\lambda_H\mu^2_H \left(\logbar{\mu^2_H} - 1\right)} \biggr. && \\
        &+ \Tilde{g}_{HR} \mu^2_R \left(\logbar{\mu^2_R}-1\right)
            + g_{HS}\mu^2_S \left(\logbar{\mu^2_S} - 1\right) && \\\nonumber
        &+ \left. a_1^2 \frac{
            \mu^2_R\left(\logbar{\mu^2_R}-1\right)-\mu^2_S\left(\logbar{\mu^2_S}-1\right)
            }{\mu^2_R-\mu^2_S} \right] \,,
\end{flalign}
\begin{flalign}\label{eq:matching_lambda}
    \nonumber
    \lambda = \lambda_H + \frac{1}{96\pi^2}\biggl[
        &\cancel{36\lambda_H^2 \logbar{\mu^2_H}} \biggr. && \\\nonumber
        &+ \Tilde{g}_{HR}^2 \logbar{\mu^2_R} + g_{HS}^2 \logbar{\mu^2_S} && \\
        &+ 9 a_1^2 \frac{
            \Tilde{g}_{HR}\left[\mu_R + \mu_S \left(\logbar{\mu_S} - \logbar{\mu_R} - 1 \right)\right]
            + R \leftrightarrow S
            }{(\mu^2_R-\mu^2_S)^2} && \\ \nonumber
        &+ \left. 6 a_1^4 \frac{
            2(\mu^2_R-\mu^2_S) - (\mu^2_R+\mu^2_S)(\logbar{\mu^2_R}-\logbar{\mu^2_S})
            }{(\mu^2_R-\mu^2_S)^3}
        \right]\ , &&
\end{flalign}
where $\logbar{m^2} \equiv \log{\frac{m^2}{\mu^2}}$ is the energy-scale-normalized logarithm\footnote{Here, $\mu$ corresponds to the \emph{hard} energy scale $\mu_{4d}$ introduced below.}, and $\Tilde{g}_{HR} \equiv g_{HR} + g'_{HR}$.
\begin{figure}[ht]
    \centering
    \includegraphics[width=\textwidth]{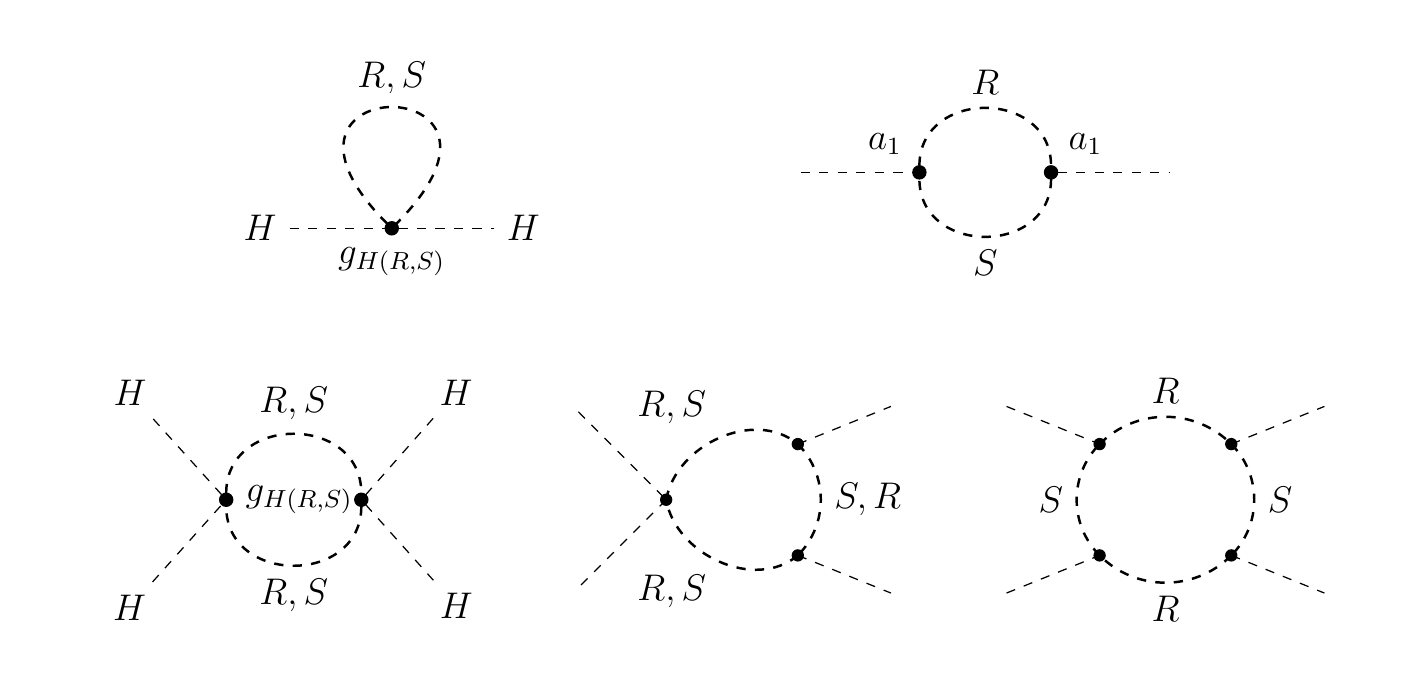}
    \caption{\footnotesize One-loop diagrams entering the SM matching Eqs.\ \labelcref{eq:matching_mu} (first line) and \labelcref{eq:matching_lambda} (second line). Thin dashed lines correspond to the Higgs field, while thick dashed lines represent one of the $R$ and $S$ LQs as indicated by the $R,S$ label.}
    \label{fig:matching_diagrams}
\end{figure}
This calculation is done by computing all diagrams that contain at least one heavy particle in the loop. This approach is equivalent to calculating all diagrams and then subtracting those that only involve light states, which cancel out when matching the two theories. These terms are crossed out in \cref{eq:matching_mu,eq:matching_lambda}. In other words, the matching conditions reflect the differences introduced by the presence of heavy particles. \cref{fig:matching_diagrams} illustrates the diagrams corresponding to the matching relations in \cref{eq:matching_mu} (first line) and in \cref{eq:matching_lambda} (second line).

We then use the SM renormalization group (RG) equations to evolve the low-energy effective parameters $\mu$ and $\lambda$ to the energy scale set by the LQ masses, approximately $\sim 1~\mathrm{TeV}$. Subsequently, we invert \cref{eq:matching_mu,eq:matching_lambda} to derive $\mu_H$ and $\lambda_H$ in terms of $\mu$, $\lambda$, and the remaining parameters of the full UV theory, denoted as $p_{LQ}$ --- specifically, $\mu_H(\mu, p_{LQ})$ and $\lambda_H(\lambda, p_{LQ})$. 

\section{Phase transitions and gravitational waves} \label{sec:PT&GW}

In this section, we outline the derivation of a finite temperature effective potential through dimensional reduction, along with the computation of phase transition parameters relevant for deriving a primordial GW spectrum. In the high-temperature EFT, we allow all three scalars in the model to acquire a VEV:
\begin{equation}\label{eq:vevConfiguration}
    H = \mqty(0 \\ v_h)\,, \qquad
    R = \mqty({\red 0} & {\blue 0} & {\green 0} \\ {\red 0} & {\blue v_r} & {\green 0})\,, \qquad
    S =  \mqty({\red v_s} & {\blue 0} & {\green 0})\,.\
\end{equation}
For simplicity, we consider only one component in the $\SU{3}{C}$ direction to acquire a VEV in both $R$ and $S$. Notice that the number of possible vacuum configurations during a phase transition increases exponentially with the number of VEVs,
\begin{equation}
    \# \text{vacuum configurations} = 2^{2\ \text{\#VEVs}}\,.
\end{equation}
This results in 64 possible configurations for our 3-VEV scenario. As we will see in \cref{sec:vac_config}, our analysis identifies around eleven configurations of interest out of all possible combinations.

\subsection{Thermodynamics of the leptoquark model}\label{sec:thermodynamics}

\begin{figure}[ht]
    \includegraphics[width=.9\textwidth]{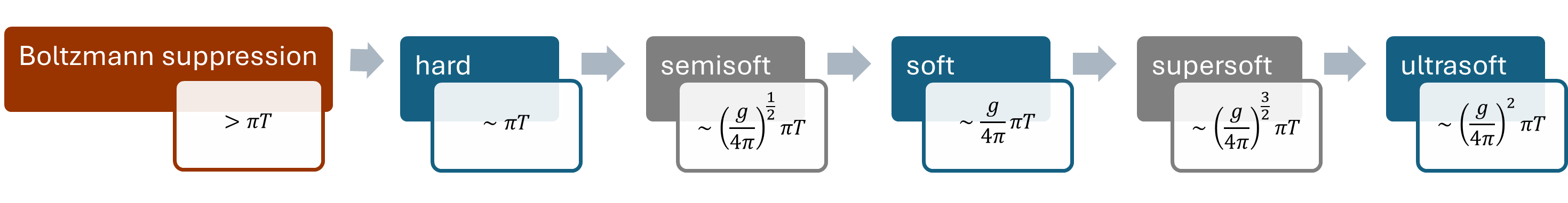}
    \caption{\footnotesize Scale hierarchies (bottom boxes) and their conventional labels (top filled boxes), where $g\ll 1$ represents a generic weak coupling and $T$ denotes the temperature. Figure inspired by Eq.\ (2.1) of \cite{Gould:2023ovu}.
    }
    \label{fig:thermalHierarchies}
\end{figure}

This study focuses on the high-temperature regime defined by
\begin{equation}
    m \ll T\ ,
\end{equation}
where $m$ represents a relevant scale associated with the magnitude of the LQ masses in our case. In this context, weakly coupled theories, such as the one under consideration, feature a hierarchy of energy scales that are determined by the temperature $T$ and the weak effective coupling $g$\footnote{In models with multiple weak couplings $g_i$, $g$ should be identified with the largest one \cite{Gould:2023ovu}. For gauge field theories, $g$ is typically identified with the gauge coupling.}. In \cref{fig:thermalHierarchies} we present these energy scales, labeled according to standard conventions and ordered from largest to smallest (from left to right). While the higher end of this hierarchy is Boltzmann suppressed\footnote{We match the thermal, 3d theory at the hard energy scale $\mu\sim\pi T$, which sets the scale of thermal fluctuations. Any phenomenon characterised by energy $E\gg \pi T$, is exponentially suppressed by a Boltzmann factor $e^{-E/T}$.}, the lighter end is non-perturbative\footnote{Each energy scale comes with its effective expansion parameter, which formally reads as $\varepsilon\sim \frac{g^2 T}{\mu} \sim 1$ (cf.\ \cite{Gould:2019qek}).}.
The 4d theory, defined at the hard scale, includes the scalar potential in \cref{eq:Vtree}, SM vector bosons, and the Yukawa Lagrangian,
\begin{equation}\label{eq:LY}
\begin{aligned}
    \mathcal{L}_\mathrm{Yukawa} =& \Delta_{ij} \Bar{Q}_i u_j \tilde{H}
        +  \Gamma_{ij} \Bar{Q}_i d_j H
        +  \Pi_{ij} \Bar{L}_i e_j H\\
        &+ \Theta_{ij} \Bar{Q}_j^c L_i S + \Omega_{ij}\Bar{L}_i d_j R^\dagger + \Upsilon_{ij} \Bar{u}_j^c e_i S + \mathrm{h.c.}\,.
\end{aligned}
\end{equation}
Here, $Q$ and $L$ are the left-handed quark and lepton $\SU{2}{L}$ doublets; $u$, $d$, and $e$ are the right-handed SM fermions. Following \cite{Freitas:2022gqs}, $\Gamma$, $\Theta$, $\Omega$, and $\Upsilon$ are generic complex $3 \times 3$ matrices, while $\Delta$ and $\Pi$ are diagonal.  $\SU{2}{L}$ contractions ($e.g.$, $\Bar{Q}^c L \equiv \epsilon_{\alpha \beta} \Bar{Q}^{c,\alpha} L^\beta$, with $\epsilon_{\alpha \beta}$ the two-dimensional Levi-Civita symbol and $c$ denoting charge conjugation) are implicit. In this article we focus on the effect of the scalar sector in producing strong FOPTs, while complying with neutrino oscillation data. Thus, we assume LQ Yukawa couplings (second line of \cref{eq:LY}) smaller than $0.01$ to safely neglect their impact on phase transitions. We have verified that for every viable FOPT, the generated LQ masses and $a_1$ trilinear coupling, there is at least one solution where the entries of the $\Theta$ and $\Omega$ matrices are small and simultaneously reproduce neutrino oscillation data. Otherwise the point is rejected. Note that an exhaustive study of flavor observables is beyond the scope if this article.

We have implemented dimensional reduction to match this 4d theory to an effective 3d theory at the \emph{soft scale} by integrating out hard-scale non-zero modes \cite{Schicho:2021gca}. Note that the EFT is three-dimensional as it solely features \emph{time-independent} static modes. For details, see \cref{Ap:DRalgo}. Additionally, we have integrated out temporal scalar fields residing at the \emph{soft} scale to derive an effective theory at the \emph{ultrasoft} scale, characterized by $\mu\sim g^2 T$. This procedure is automated in \texttt{DRalgo} \cite{Ekstedt:2022bff}, from where we obtained our dimensionally reduced effective potential at next-to-leading-order (NLO). We perform the matching at the energy scale $\mu=\pi T$. The resulting 3d effective potential takes the form $V_{\text{eff}} = V_{\text{eff}}^{(0)} + V_{\text{eff}}^{(1)}$, with
\begin{equation}\label{eq:Veffective}
\begin{aligned}
    V_{\text{eff}}^{(0)} =& \quad\ \frac{1}{2}\qty(\hat\mu^2_H v_h^2  + \hat\mu^2_R {\red v_r}^2  + \hat\mu^2_S {\blue v_s}^2) \\
        & + \left[\hat\lambda_H v_h^4 + \hat\lambda_R {\red v_r}^4 + \hat\lambda_S {\blue v_s}^4 \right.\\
        & \qquad\quad\ +\left. \hat g_{HS} v_h^2 {\blue v_s}^2 + (\hat g_{HR} + \hat g'_{HR}) v_h^2 {\red v_r}^2 + \hat g_{RS} {\red v_r}^2 {\blue v_s}^2\right]  \\
        & + \hat a_1 v_h {\blue v_s} {\red v_r} \\
    V_{\text{eff}}^{(1)} =& -\frac{1}{12\pi} 
    \left(\sum_i m_{S,i}^3 + \sum_i m_{V,i}^3 \right) \,.
\end{aligned}
\end{equation}
Here, $m_S$ and $m_V$ represent the scalar and vector boson masses in the \emph{ultrasoft} EFT. All model parameters above, denoted with a \emph{hat}, are calculated at the \emph{ultrasoft} scale according to \cref{Ap:DRalgo}. These parameters are matched to the 4d potential in \cref{eq:Vtree} and depend implicitly on the temperature, with the following substitutions applied to the 3d effective potential above:
\begin{align}\label{eq:V3_2_V4}
    v_i^\mathrm{3d} &\rightarrow \frac{v_i^\mathrm{4d}}{\sqrt{T}} \\
    V_{\text{eff}}^\mathrm{3d} &\rightarrow T V_{\text{eff}}^\mathrm{3d} \equiv V_{\text{eff}}^{\mathrm{4d}}\,.
\end{align}
Here, $v_i$ represents the VEVs in \cref{eq:vevConfiguration}. This procedure rewrites the 3d fields and potential in the form of a 4d theory \cite{Ekstedt:2022bff}.

\subsection{Features of a cosmological phase transition} \label{sec:phaseTransitions}

Phase transition parameters are conventionally computed at a fixed temperature $T_*$, which, following recent literature \cite{Athron:2023xlk}, is chosen to be the percolation temperature $i.e.$, $T_*\equiv T_p$\footnote{This choice is more appropriate for the transition temperature than the nucleation temperature, as it remains valid in strongly supercooled scenarios. See \cite{Athron:2023xlk} for further discussion.}. In this context, the key quantity to consider is the false vacuum fractional volume $P_{\mathcal{F}}(T) \equiv e^{-I(T)}$, with $I(T)$ defined by
\begin{equation} \label{eq:frac_vol_log}
    \mathcal{I}(T) = \frac{4\pi}{3}\xi_w^3\int^{T_c}_{T}\dd{T'}\frac{\Gamma(T')}{H(T')}\left(\int^{T'}_{T}\frac{dT''}{H(T'')}\right)^{3}\ .
\end{equation}
In the expression above, $\xi_w$ is the bubble wall velocity (in natural units), $H(T)$ is the Hubble parameter, and 
\begin{equation} \label{eq:decayRate}
    \Gamma(T) = T^4 \left(\frac{S_3}{2\pi T}\right)^\frac{3}{2} e^{-\frac{S_3}{T}}
\end{equation}
is the false vacuum decay rate density for thermal phase transitions\footnote{The prefactor in \cref{eq:decayRate} arises from a dimensional estimate of the bounce action functional determinants at high temperature, which are notoriously difficult to compute.} \cite{LINDE1983421}. Assuming $O(3)$ symmetry around a center of the bounce, the 3d Euclidean bounce action can be written as \cite{LINDE1983421}
\begin{equation}
    S_3 = 4\pi\int \dd{\rho} \rho^2 \left[\frac{1}{2}\left(\frac{d\phi}{d\rho}\right)^2 + V(\phi)\right]\ .
\end{equation}
Although the model predominantly features phase transitions with mild supercooling, as discussed in the context of \cref{fig:supercooling}, we include the vacuum contribution to the total energy density entering the Hubble parameter $H^2 = \frac{\rho}{3M_P^2}$ \cite{Hindmarsh:2020hop,Athron:2023xlk}:
\begin{equation} \label{eq:rhoEnergyDensity}
    \rho(T) = \underbrace{\frac{\pi^2}{30} g_* T^4}_{\text{radiation}} + 
        \underbrace{\vphantom{\frac{\pi^2}{30}}\Delta V}_{\text{vacuum}}\ ,
\end{equation}
where $g_*$ is the effective number of relativistic degrees of freedom at $T_*$. This approach allows us to account for the few scenarios with strong supercooling identified in our numerical analysis where $\Delta V$ dominates the energy density of the Universe.

The nucleation temperature $T_n$ is generically defined by
\begin{equation} \label{eq:nucleationFull}
    \int_{T_n}^{T_c} \dd{T} \frac{\Gamma(T) P_\mathcal{F}(T)}{T H(T)^4} = 1\ .
\end{equation}
Here, $P_{\mathcal{F}}(T>T_n)$ denotes the probability of being in the false vacuum prior to nucleation, which is typically the case\footnote{This is usual except for very strongly supercooled phase transitions, where a few bubbles may nucleate and expand to occupy a large fraction of the Universe's volume before reaching the nucleation condition defined in \cref{eq:nucleationFull}.}, $i.e.$ $P_{\mathcal{F}}(T>T_n) \approx 1$, from which the usual expression is derived
\begin{equation} \label{eq:nucleationIntGammaOverH4}
    \int_{T_n}^{T_c} \dd{T} \frac{\Gamma(T)}{T H(T)^4} = 1\ .
\end{equation}
In numerical calculations, it is useful to provide an initial estimate for the nucleation temperature using the condition:
\begin{equation} \label{eq:nucleationGammaOverH4}
    \frac{\Gamma(T_n)}{H^4(T_n)} = 1\ .
\end{equation}
The most commonly adopted nucleation criterion for EW-scale phase transitions is based on the Euclidean action value \cite{Caprini_2020} which is approximately given by $S_3/T\sim140$. However, the LQ model features phase transitions over a range of energies from $100~\mathrm{GeV}$ to approximately $10~\mathrm{TeV}$, corresponding to $S_3/T$ values at nucleation of roughly $(137-124)$. In our numerical routines, we use \cref{eq:nucleationGammaOverH4} to identify efficient FOPTs and later refine the nucleation temperature determination using \cref{eq:nucleationIntGammaOverH4}. For further details, see \cref{sec:numerics}.

\begin{figure}[!ht]
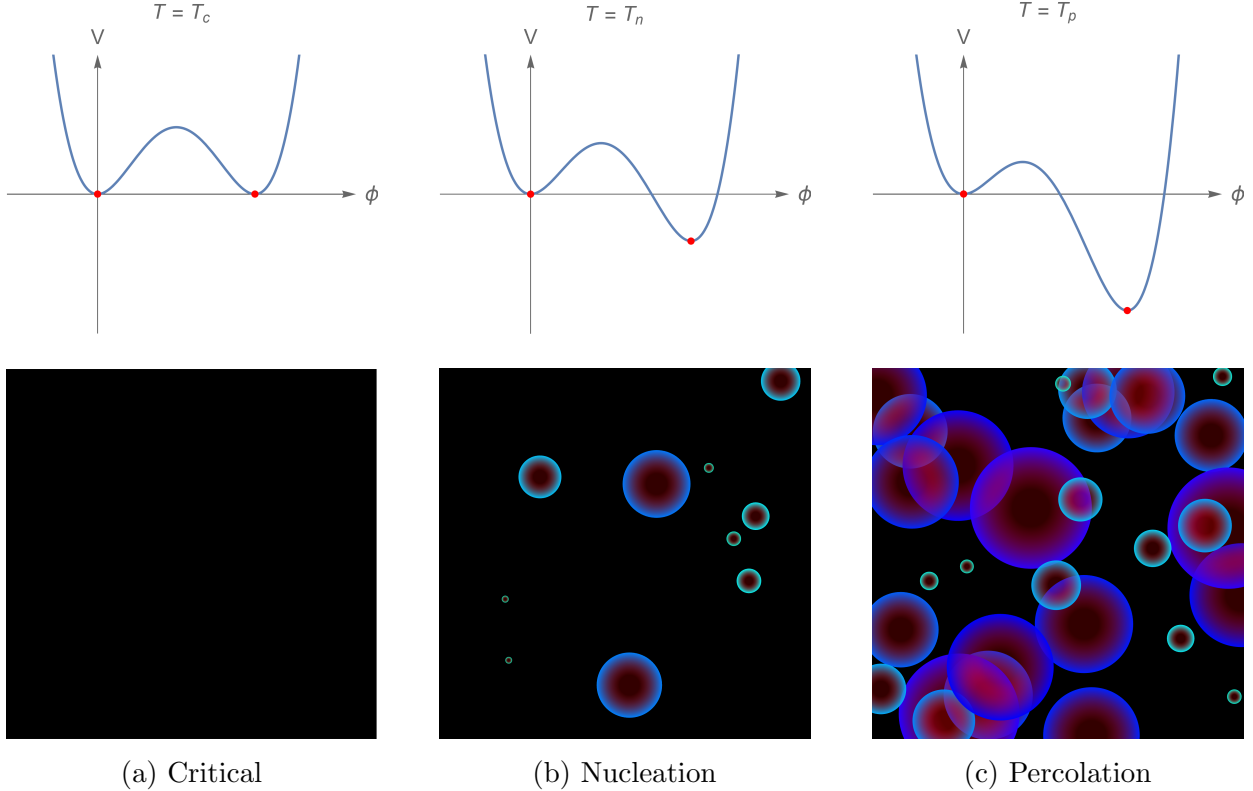

    \include{figs/Graphics/transitionTemperatures}
    \caption{\footnotesize Schematic representation of a FOPT, characterized by three temperatures. From left to right: At the critical temperature $(T_c)$ the minima are degenerate, quantum tunneling to the new phase is not favoured and the Universe remains entirely in the original phase. At the nucleation temperature $(T_n)$, the transition to the new phase becomes sufficiently energetically favored, allowing for the efficient nucleation of true vacuum bubbles. Finally, at the percolation temperature $(T_p)$, the bubbles form a causally connected structure throughout the Universe, preventing a collapse back into the false vacuum.
    }
    \label{fig:transition_temperatures}
\end{figure}

The percolation temperature can be determined by setting the fractional false-vacuum volume to $P_{\mathcal{F}}(T) \approx 0.71$\footnote{The estimate arises from percolation analyses of uniformly nucleated spherical bubbles \cite{Athron:2023xlk}.}, or equivalently
\begin{equation} \label{eq:percolation}
    \mathcal{I}(T) \approx 0.34\,.
\end{equation}
\cref{fig:transition_temperatures} schematically illustrates the three stages that characterize a phase transition, from which we derive the critical, nucleation, and percolation temperatures (as shown in panels (a) to (c), respectively). Since the Euclidean bounce action must be computed numerically at fixed temperatures, fulfilling this condition is computationally prohibitive. Consequently, we fit the action over the range of temperatures of interest, as described in \cref{sec:numerics}.

Taking into account the expansion rate of the Universe, it is essential to ensure that the true vacuum volume $\mathcal{V}(t)=a^3(t)P(t)$ is increasing at the time of percolation \cite{Athron:2023xlk}. From $\mathcal{V}'(t)\ge0$, and the time-temperature relation $\frac{dT}{dt}=-T H(T)$, one can directly derive 
\begin{equation}
    \eval{H(T)(T \mathcal{I}'(T) + 3)}_{T_p} < 0\ .
\end{equation}
This condition can, in principle, be violated, preventing true percolation from being achieved. This applies in particular to strongly supercooled transitions.

Besides temperature, the thermodynamics of a phase transition is fully described by two quantities that represent the available energy and the characteristic time scale \cite{Hindmarsh:2020hop,Athron:2023xlk}. These are captured by the transition strength  
\begin{equation}\label{eq:alpha}
    \alpha = \eval{\frac{1}{\rho_\gamma}\left[\Delta V - \frac{T}{4}\Delta\left(\frac{\partial V}{\partial T}\right)\right]}_{T_*} \,,
\end{equation}
and the inverse duration in Hubble units
\begin{equation}
    \frac{\beta}{H} = T_{*}\frac{\partial}{\partial T}\left.\left(\frac{S_3}{T}\right)\right\vert_{T_{*}}\ ,
\end{equation}
respectively. A typical choice for the characteristic length scale is the mean bubble separation, which is derived from the bubble number density $n$ and estimated by $\beta$:
\begin{align}
    R_* &= n(T_*)^{-1/3} \\
        &=  T_*^3 \int_{T_*}^{T_c} \dd{T} \frac{\Gamma(T)P_\mathcal{F}(T)}{T^4 H(T)} \\
        &\approx (8\pi)^{1/3}\frac{\max(\xi_w,c_s)}{\beta} \,,
\end{align}
where $c_s$ is the speed of sound in the false vacuum \footnote{For an ultra-relativistic plasma, the speed of sound in both phases is given by $c_s^2=1/3$.}. The approximation in the last line holds for fast phase transitions.

\subsection{Gravitational wave observables}

GWs from FOPTs can be sourced through three distinct channels \cite{Caprini:2024hue}:
\begin{outline}
    \1 bubble wall collisions;
    \1 sound waves;
    \1 Magnetohydrodynamic (MHD) turbulence.
\end{outline}
Hydrodynamic simulations indicate that, except in cases of extremely strong transition, $i.e.$ $\alpha \gg 1$, the contribution from bubble wall collisions is typically subdominant \cite{Caprini:2018mtu,Ellis:2019oqb}. Since our analysis focuses only on GW spectral peaks, where the contribution from MHD turbulence is also generally subdominant, we will concentrate on GWs sourced by sound waves. Indeed, for most scenarios considered here, we find that $\alpha < 1$ with a few exceptions where it can reach up to $\mathcal{O}(10)$.

We adopt the GW templates from Ref.\ \cite{Caprini:2024hue}, which provides a broken power law for the GW spectra sourced by bubble collisions and a doubly broken power law for sound waves and turbulence. The frequency breaks for sound waves are
\begin{align}
    f_1 &\simeq 0.2 H_{\text{reh},0} (H_*R_*)^{-1} \,, \\
    f_2 &\simeq 0.5 H_{\text{reh},0} (H_*R_*)^{-1} \Delta_w^{-1} \,,
\end{align}
where $\Delta_w=\frac{\abs{\xi_w-c_s}}{\max(\xi_w,c_s)}$ represents the ratio of the sound shell thickness to the wall velocity (or the speed of sound in case of subsonic deflagrations). We compute the Hubble rate at the time of GW production $H_\text{reh}$, redshifted to the modern Universe, as follows:
\begin{equation}
    H_{\text{reh},0} \simeq 1.65\times10^{-5} \left(\frac{g_*}{100}\right)^{1/6} \left(\frac{T_\text{reh}}{100 \ \text{GeV}}\right) \ \text{Hz}\,.
\end{equation}
This is evaluated at the reheating temperature \cite{Ellis:2018mja}
\begin{equation}
    T_{\text{reh}} \simeq T_p (1+\alpha)^{1/4}\ ,
\end{equation}
considering that immediately after percolation, the latent heat stored in the false vacuum is released and reheats the Universe. While this effect is primarily relevant when $\alpha \gg 1$, typically resulting from strongly supercooled FOPTs, we include it in our calculation for completeness and to account for the few strong phase transitions identified in \cref{fig:supercooling}.

The sound wave energy density can be decomposed as
\begin{equation}
    h^2\Omega_{\rm GW} = h^2\Omega_2 \frac{S(f)}{S(f_2)}\ ,
\end{equation}
where the spectral shape captures the double broken power law
\begin{equation}
    S(f) = \frac{N}{f_1^3} \frac{f^3}{\left(1+\frac{f}{f_1}\right)^2 \left(1+\frac{f}{f_2}\right)^4}\ ,
\end{equation}
and the GW amplitude is given by
\begin{equation}
    \Omega_2 = \frac{1}{\pi}\left(\sqrt{2}+\frac{2f_2/f_1}{1+(f_2/f_1)^2}\right) F_{{\rm GW},0} A_{\text{sw}} K^2 (H_*\tau_{\text{sw}}) (H_*R_*)\ .
\end{equation}
In this expression, $h^2F_{{\rm GW},0}\simeq 1.64\times10^{.-5} (100/g_*)^{1/3}$ is the redshift factor for the fractional energy density $K\simeq 0.6\kappa\alpha/(1+\alpha)$, and $A_{\text{sw}}\simeq 0.11$ is a constant characteristic of sound waves. The relative duration of the GW source is quantified by $H_*\tau = \min(H_*R_*/\sqrt{3K/4},1)$. We use the semi-analytical fit functions derived in \cite{Espinosa_2010} for the efficiency coefficient $\kappa$\footnote{The relations hold for a perfect ultra-relativistic fluid with a speed of sound given by $c_s^2 = 1/3$.}.

We then maximize this spectrum to determine the GW peak frequency and amplitude. These values are compared to the \emph{peak-integrated sensitivity curves} (PISC) of Ref.\ \cite{Schmitz:2020syl} for the LISA \cite{Kawamura:2006up}, DECIGO and BBO \cite{Harry:2006fi} detectors. The vertical distance of each peak to a given curve indicates the signal-to-noise ratio for the corresponding detector. 

\section{Numerical analysis} \label{sec:numerics}

Our numerical analysis employs the thermal effective potential derived from the scalar sector potential in \cref{eq:Vtree} via dimensional reduction using the \texttt{DRalgo} tool \cite{Ekstedt:2022bff} (\cref{sec:thermodynamics,Ap:DRalgo}). The model and the effective potential are implemented using \texttt{Dratopi} \cite{Dratopi},
a soon-to-be-released tool for phase transition analysis in the dimensional reduction formalism. The tool interfaces \texttt{DRalgo} with \texttt{Python} and a slightly modified version of \texttt{CosmoTransitions} \cite{Wainwright_2012}, to facilitate phase transition calculations for generic models in \texttt{Python} using the dimensional reduction approach. The package includes a \texttt{Mathematica} script to automate the export of models constructed within \texttt{DRalgo} to \texttt{Python}. \texttt{Dratopi} then implements several protocols within a modified version of \texttt{CosmoTransitions}, including:
\begin{outline}
    \1 Numerically solving the RG equations in the 4d theory, with the possibility of specifying constraints, such as perturbativity constraints, on the parameters in 4d theory.
    \1 Automatically determining minimal constraints on the temperature range of the model (required by the positivity of the squared Debye masses).
    \1 Implementing the dimensional reduction (at LO or NLO), through the expressions exported from \texttt{DRalgo}, to construct the 3d EFT in the ultrasoft regime.
    \1 Implementing the effective potential (at LO, NLO or NNLO), through the expressions exported from \texttt{DRalgo}. In particular, this includes an automated procedure to (numerically) diagonalize the mass matrix in an efficient manner, as needed for the NLO and NNLO parts of the effective potential.
    \1 Tracing the phases and searching for phase transitions, using slightly tweaked versions of the \texttt{CosmoTransitions} routines, but starting at a non-zero temperature, as necessary in the dimensional reduction approach.
    \1 Providing various sanity checks, notably the ratio $m_{\text{US}}(\phi,\mu)/\mu$ of the effective mass to energy scale, at a given energy scale and field values (see \cref{sec:high-T}).
\end{outline}
\cref{Ap:DRalgo} provides further details on dimensional reduction and \texttt{Dratopi}. Documentation for the package will also be publicly available shortly. In what follows we present some important observations concerning our numerical implementation of the LQ model. Note that, for brevity, we will typically use the notation $\mu$ to refer to the hard energy scale, or hard matching scale, $\mu_{4d}$, discussed in more detail in \cref{Ap:DRalgo}.

\paragraph*{Relations to LQ masses --}
Collider experiments establish a lower bound on scalar LQ masses of approximately $m_{LQ}^{\text{min}} \sim \mathcal{O}(1~\mathrm{TeV})$. In our numerical analysis, we fix the mixing angle $\theta$ and the LQ mass ($\mS{1,2}$), ensuring it satisfies this constraint, and then determine $\mS1$, $\ms$, $\mu_S$, and $\mu_R$ by inverting the relations \labelcref{eq:lq_masses,eq:mixingAngle}:
\begin{align}\label{eq:lq_masses_invert}
    \mu_R^2 &= \frac{1}{2} \left(\mS1^2+\mS2^2 - g_{HS}v_H^2
        + m_c^2\right)\,, \\
    \mu_S^2 &= \frac{1}{2} \left(\mS1^2+\mS2^2 - (g_{HS}+g_{HR}')v_H^2
        - m_c^2\right)\,, \\
    \mS1^2 &= \mS2^2 - \frac{\sqrt{2}v_H a_1}{\sin(2\theta)}\,, \\
    \ms^2 &= \mu_R^2 + \frac{1}{2}g_{HR}v_H^2\,,
\end{align}
where we define
\begin{equation}
    m_c^4 \equiv \left(\mS2^2-\mS1^2\right)^2 - 2 a_1^2 v_H^2\ .
\end{equation}

The mass-matrix parameterization using the mixing angle $\theta$ yields \cref{eq:mixingAngle}. Given that $|\sin(2\theta)| < 1$ and $m_{LQ}^{\text{min}} < \mS1 < \mS2$, this implies an upper bound on the trilinear coupling $a_1$, which we required to be positive:
\begin{equation} \label{eq:a1Max}
    a_1 < a_1^{\text{max}} \equiv \frac{\mS2^2-(m_{LQ}^{\text{min}})^2}{\sqrt{2}v_H}\ .
\end{equation}

\paragraph*{Parameter ranges --}
The ranges of parameters used in our scans are shown in \cref{tab:scan_parameters}.
\begin{table}[!ht]
    \centering
    \begin{tabular}{|c|c|} \hline 
        Parameter & Range\\ \hline \hline
        $\log_{10}\lambda_{S,R}$ & $\left(-3, \log_{10}{2}\right)$ \\ \hline 
        $\log_{10}g$ & $\pm \left(-3,  \log_{10}{2}\right)$ \\ \hline 
        ${\mS2^2}_{\vphantom{a_b}}$ & $\left(0.8,3\right)$ TeV \\ \hline
        $\log_{10}a_1$& $\left(-2,\log_{10}{a_1^{\text{max}}}\right)$ TeV \\ \hline 
        $\theta$& $\left(0,\pi/2\right)$ \\ \hline
    \end{tabular}
    \caption{\footnotesize Free-parameter ranges for the scanning routines. Here, $g$ stands for any of the mixed quartic couplings $\{g_{HS},g_{HR},g_{HR}',g_{RS}\}$. Since these can be negative, we introduce a (uniformly) random sign $\pm$.}
    \label{tab:scan_parameters}
\end{table}

Unitarity and perturbativity constraints limit the quartic couplings to values below $4\pi$ and approximately $\mathcal{O}(1)$, respectively. Here, we allow values up to 2. To ensure a bounded-from-below tree-level potential (requiring positive self-couplings $\lambda$), these constraints are enforced across the entire energy range of our scans using the phase-tracing functionality of \texttt{Dratopi}. Phase tracing terminates if any constraint is violated. \cref{eq:a1Max} provides an upper bound for the trilinear coupling $a_1$. The lower bound, $10^{-2}$, is significantly smaller than the LQ masses and can be considered negligible. As discussed in \cref{sec:LQmasses}, requiring $a_1 > 0$ restricts the allowed range of the mixing angle to the interval $(0, \pi/2)$.

\paragraph*{Action fitting --}
To determine the thermodynamic parameters detailed in \cref{sec:thermodynamics}, we must compute the effective tunneling action. In the case of thermal phase transitions, this involves solving the three-dimensional bounce equation \cite{LINDE1983421,Athron:2023xlk}, given by
\begin{equation} \label{eq:bounce}
    \frac{d^2\mathbf{\phi}}{d\rho^2} + \frac{2}{\rho}\frac{d\mathbf{\phi}}{d\rho} = \nabla_\mathbf{\phi} V_\text{eff}(\phi,T)\,.
\end{equation}
This equation is solved with the boundary conditions
\begin{align} \label{eq:bounceBoundCond}
    \mathbf{\phi}(\rho\rightarrow 0) &= \mathbf{\phi}_\mathcal{F} \\
    \eval{\frac{d\mathbf{\phi}}{d\rho}}_{\rho=0} &= 0\ ,
\end{align}
where $\mathbf{\phi}_\mathcal{F}$ corresponds to the field value in the false vacuum. 

As noted in \cref{sec:phaseTransitions}, the computation of the thermal tunneling action using the methods implemented in \texttt{CosmoTransitions} is susceptible to numerical instabilities \cite{Freitas:2021yng}. This is further compounded by the computationally expensive nature of directly determining the percolation temperature from \cref{eq:percolation}, which requires solving the bounce equation numerically for a large number of temperatures. To address these computational challenges, we adopt a more efficient approach. Specifically, we perform a numerical fit of the bounce action around the nucleation temperature, which is itself estimated from the nucleation criterion given in \cref{eq:nucleationGammaOverH4}. This fit employs a Laurent polynomial of the form,
\begin{equation}
    \frac{S_3}{T} \approx \sum_{n=-3}^3 c_n\left(T-T_c\right)^n \,,
\end{equation}
to capture the bounce action's behavior near the nucleation temperature (positive powers) and its divergence at the critical temperature (negative powers). This analytical expression for the bounce action as a function of temperature allows for the numerical evaluation of the integrals in \cref{eq:nucleationGammaOverH4,eq:percolation}. This procedure yields an improved estimate of the percolation temperature and refines the calculation of the nucleation temperature.

\subsection{Gravitational wave peak amplitudes} \label{sec:GWpeaks}

We performed a comprehensive scan of the LQ model's parameter space, employing the templates in \cite{Caprini:2024hue} to extract the relevant phase transition and GW peak-amplitude parameters. This resulted in the generation of approximately $\mathcal{O}(10^5)$ points featuring the phase transitions. For a generic BSM theory to faithfully represent a SM-like 3d EFT, a specific region in the $\alpha-\beta/H$ parameter space was identified in \cite{Gould:2019qek}. Transitions exhibiting weaker strength (smaller $\alpha$) and shorter duration (larger $\beta/H$) than those within that parameter space region are effectively crossovers.  To simplify the analysis, we impose the constraint $\beta/H \lesssim 10^5$. It is important to note that points violating this constraint are, in all cases, well outside the region of observational sensitivity. Approximately $2 \times 10^4$ data points satisfy this criterion and are highlighted in color in the following GW peak-amplitude distributions.

Prior to presenting our results, we discuss the treatment of the bubble wall velocity, $\xi_w$. A rigorous determination of $\xi_w$ would require computationally intensive lattice simulations to account for the effects of plasma friction on the expanding bubbles. Since such simulations are beyond the scope of this work, we adopt a simplified approach.  We assume that the bubble walls expand via supersonic detonations and fix the bubble wall velocity to $\xi_w = 0.95$ throughout our analysis. This approximation is justified by the relatively strong nature of the phase transitions considered, specifically those within reach of planned GW detectors. These transitions consistently exhibit values of $\alpha > 10^{-2}$, which, as demonstrated in \cite{Laurent:2022jrs}, correspond reasonably well to the supersonic detonation regime characterized by $\xi_w \approx \mathcal{O}(1)$.

\begin{figure}[!htb]
\begin{subfigure}{.5\textwidth}
  \centering
  \includegraphics[width=1.\linewidth]{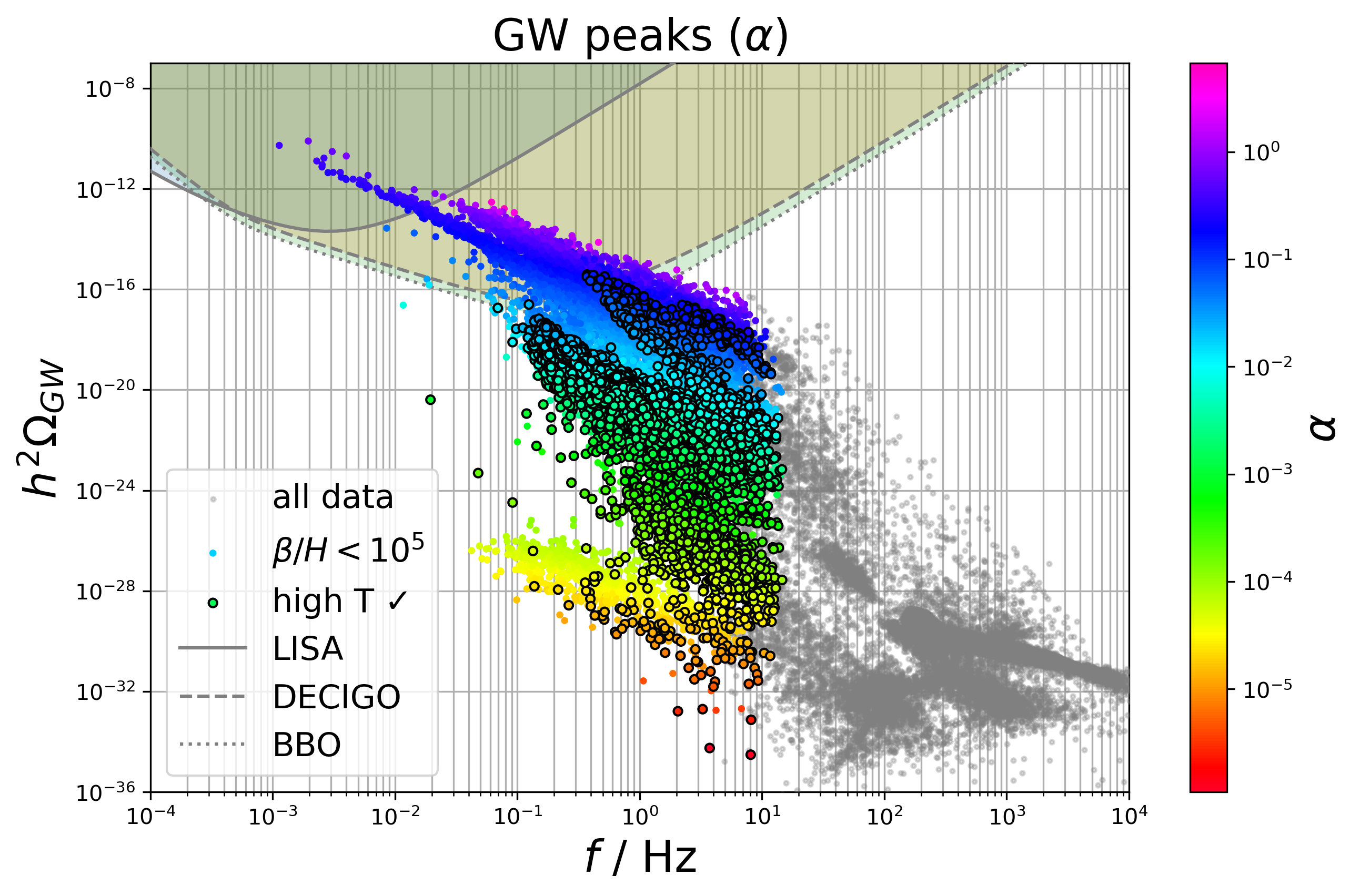}
\end{subfigure}%
\begin{subfigure}{.5\textwidth}
  \centering
  \includegraphics[width=1.\linewidth]{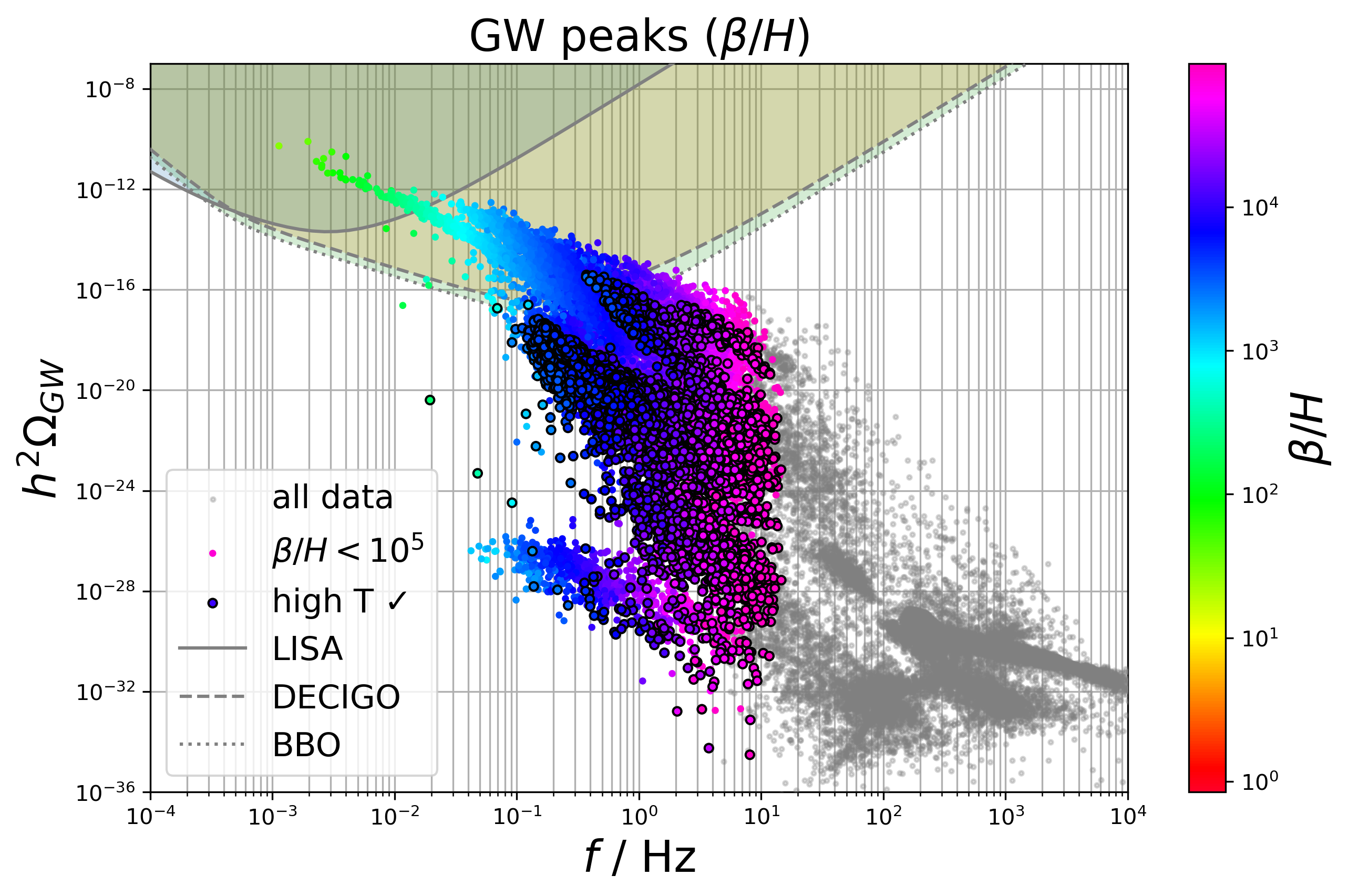}
\end{subfigure}%
    \caption{\footnotesize GW spectral peaks of the combined numerical scans. Gray points correspond to phase transitions occurring too quickly for the EFT treatment of \cref{sec:thermodynamics} \cite{Gould:2019qek}.
    Color represents the strength $\alpha$ of the phase transitions (left panel) and inverse duration $\beta/H$ (right panel). Circled dots highlight the phase transitions \emph{strictly} consistent with the high-$T$ expansion, see \cref{sec:high-T}. The PISCs for LISA, DECIGO and BBO are reported for comparison.}
    \label{fig:GWpeaks}
\end{figure}

\cref{fig:GWpeaks} presents the logarithmic distribution of the GW peak amplitude, $h^2\Omega_{\rm GW}^{\text{peak}}$, as a function of the peak frequency and the released latent heat (color scale on the left panel) and the inverse duration of the phase transition (color scale on the right panel). Inspecting this distribution reveals the minimum transition strength and the maximum duration required to achieve detectability with different GW detectors. Specifically, a transition strength of $\alpha \gtrsim 10^{-1}$ and $\beta/H \lesssim 10^2$ is necessary to approach LISA sensitivity, whereas a strength of $\alpha \gtrsim 10^{-2}$ and $\beta/H \lesssim 10^4$ is required to reach DECIGO and BBO sensitivities.

\subsection{High-temperature perturbativity} \label{sec:high-T}

The thermal effective potential calculated using \texttt{DRalgo} and given in \cref{eq:Veffective} is valid at high temperatures. However, to maintain control over infrared (IR) logarithmic corrections within the dimensional reduction framework, it is crucial to satisfy the condition $m_{\text{US}}(\phi, \mu) < \mu$, where $m_{\text{US}}(\phi, \mu)$ represents the various bosonic masses in the ultrasoft effective theory, and $\mu$ is the hard energy scale \cite{Kajantie:1995dw,Ekstedt:2022bff,Kierkla:2023von}.

\begin{figure}[!htb]
    \centering
    \includegraphics[width=0.5\linewidth]{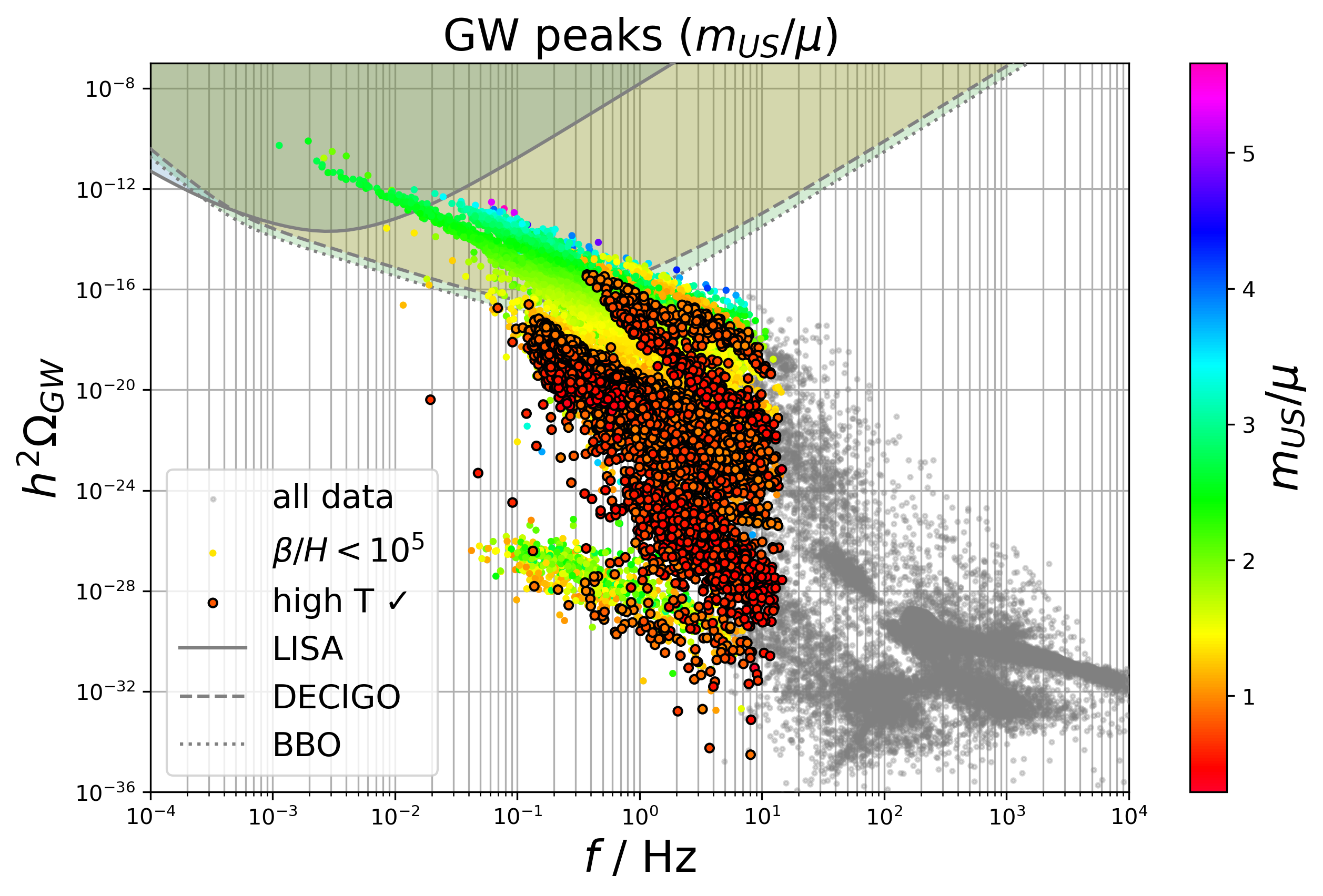}
    \caption{\footnotesize Distribution of GW peak-amplitudes color-coded according to the high-temperature perturbativity parameter $\eval{m_{\text{US}}/\mu}{T_n}$, where $m_{\text{US}}$ is the maximum ultrasoft mass at the nucleation temperature.  Points satisfying the perturbativity criterion ($m_{\text{US}}/\mu < 1$) are circled.
    }
    \label{fig:highTcheck}
\end{figure}

In this work, we distinguish between phase transitions that strictly satisfy the high-temperature perturbativity constraint given in \cref{eq:highTcheck} and those that do not. Transitions satisfying this constraint are identified in \cref{fig:GWpeaks,fig:highTcheck} using circled dots and are labeled as ``high $T$ $\checkmark$''. The validity of this constraint is verified at the nucleation temperature by calculating the ratio of the highest ultrasoft mass to the energy scale as follows:
\begin{equation} \label{eq:highTcheck}
    \eval{\frac{m_{\text{US}}}{\mu}}_{T_n} < 1 \quad \forall m_{\text{US}} \,.
\end{equation}
Given that the effective masses depend on the field values, we determine the maximum effective mass at both the initial and final phase VEVs characterizing the phase transition. This maximum value is then used to assess the validity of the high-temperature approximation.  The resulting distribution of GW peak-amplitudes is shown in \cref{fig:highTcheck}; this figure is identical to \cref{fig:GWpeaks} except that the data points are now color-coded according to the ratio $m_{\text{US}}/\mu$.

\begin{figure}[!ht]
\begin{subfigure}{.5\textwidth}
  \centering
  \includegraphics[width=1.\linewidth]{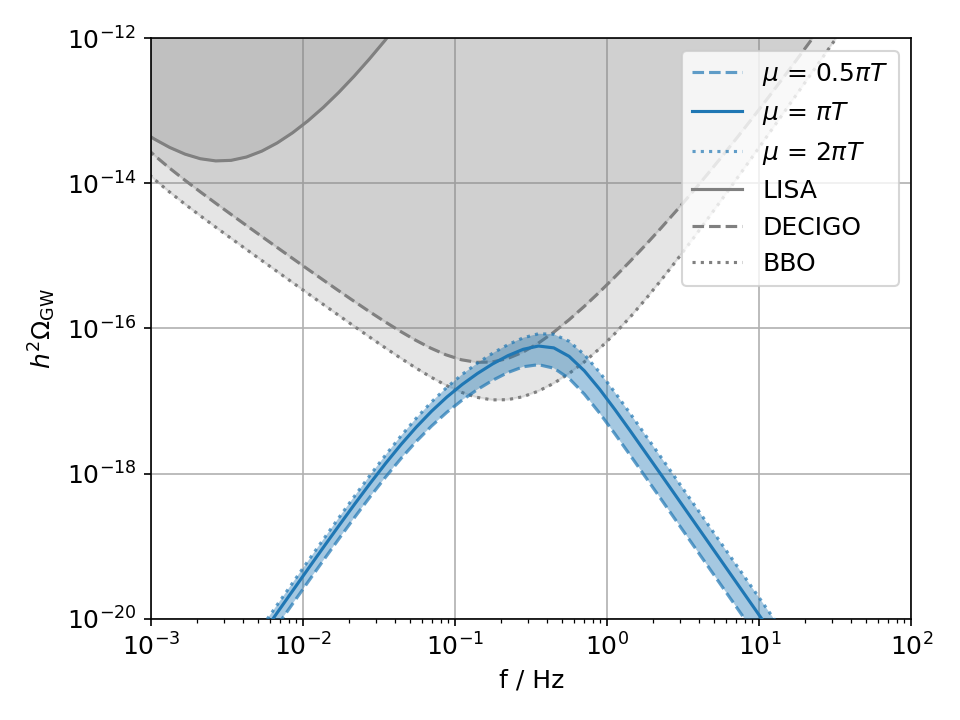}
  \caption{BP1 with $m_{\text{US}}/\mu\approx 0.75$.}
\end{subfigure}%
\begin{subfigure}{.5\textwidth}
  \centering
  \includegraphics[width=1.\linewidth]{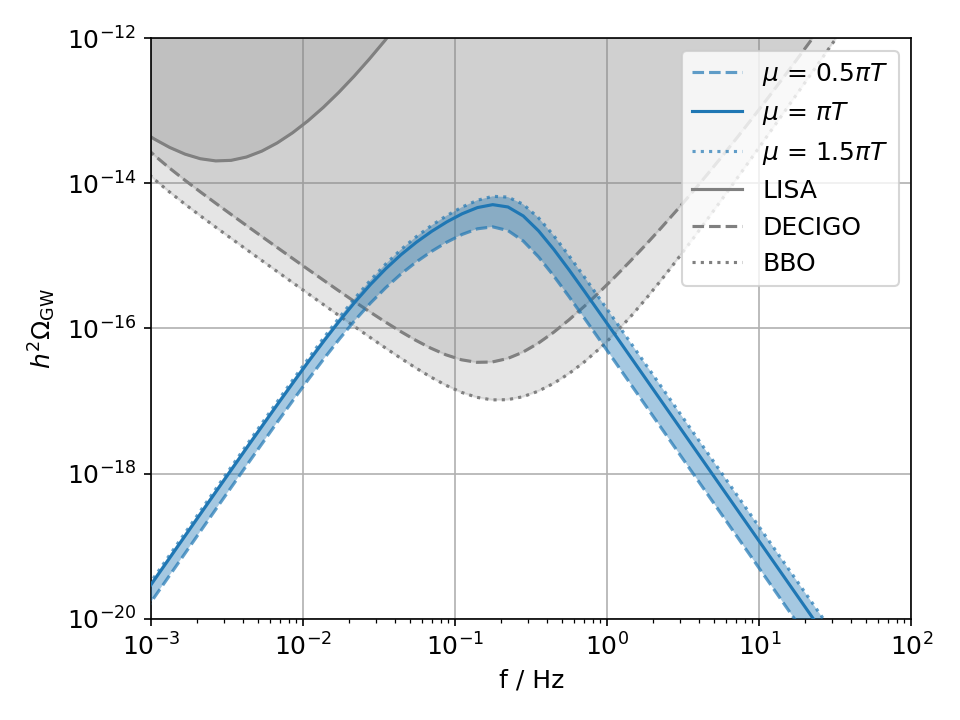}
  \caption{BP2 with $m_{\text{US}}/\mu\approx 2$.}
\end{subfigure}%
    \caption{\footnotesize GW predictions for two benchmark points (BP) within the LQ model parameter space, characterized by a color restoration $r \to h$ transition. The blue shaded regions represent the uncertainties associated with variations in the energy-scale factor $\mu/(\pi T)$. The narrowness of these uncertainty bands provides strong evidence supporting the validity of the high-temperature perturbative approximation employed in this analysis.}
    \label{fig:GWspectra_scaleFactor}
\end{figure}

The high-temperature perturbativity condition specified in \cref{eq:highTcheck} is satisfied to a good approximation for all sampled points in our analysis. For points with lower GW amplitudes ($h^2 \Omega_\mathrm{\rm GW} \lesssim 10^{-15}$), the condition is satisfied in most of the cases. However, it is crucial to understand that this condition does not represent a strict upper limit on the allowed values. Instead, it should be interpreted as an approximate measure of the overall perturbativity of the 3d EFT at high temperatures. For a finer analysis higher-order corrections on the thermal effective potential are needed, which is beyond the scope of this article. In this context, deviations by a factor of a few above unity are considered acceptable. Interestingly, even the strongest transitions, including some within the sensitivity range of LISA, exhibit values of the mass-to-energy scale ratio ($m_{\text{US}}/\mu$) that deviate slightly from 1, with the majority of points clustering around a value of 2. The robustness of the high-temperature approximation is further assessed by varying the energy-scale factor $\mu/(\pi T)$. The impact of this variation on the predicted GW spectra is illustrated in \cref{fig:GWspectra_scaleFactor} for two benchmark points (BP) describing transitions involving the color restoration pattern $r\rightarrow h$. This figure demonstrates that varying the energy-scale factor $\mu/(\pi T)$ within the range [0.5, 2] produces only a small change in GW spectral amplitudes, $\Delta h^2\Omega_{\text{GW}} < 0.5$. This highlights the significant reduction in theoretical uncertainties achieved through dimensional reduction, compared to standard 4d calculations.
\begin{figure}[!ht]
    \centering
    \includegraphics[width=.87\linewidth]{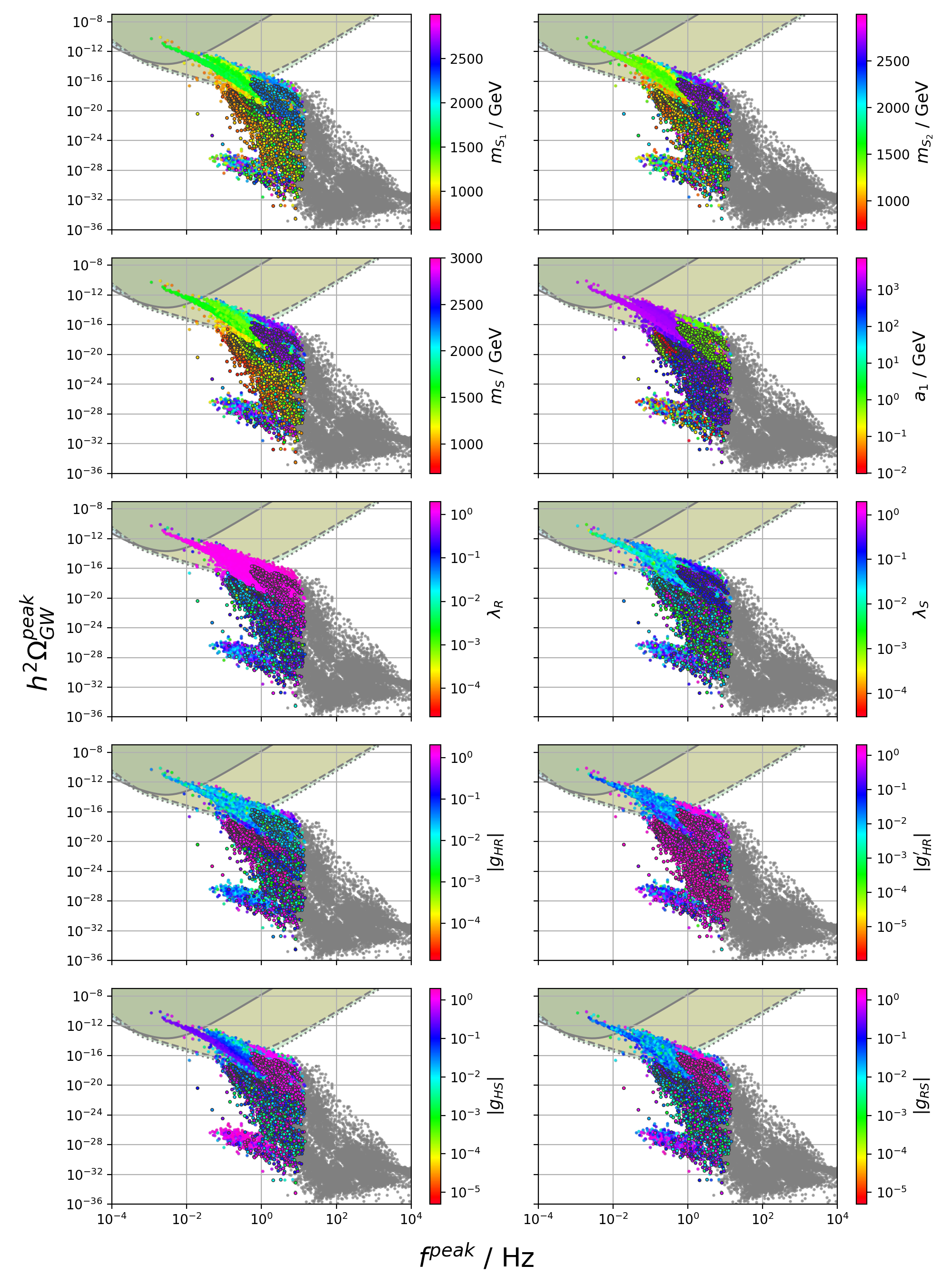}
    \caption{\footnotesize Predictions for the GW peak amplitude and frequency for various model parameters. The plots are organized as follows: Row 1 shows the dependence on two of the physical LQ mass parameters. Row 2 shows the dependence on the third LQ mass parameter (left panel) and the trilinear coupling $a_1$ (right panel). Row 3 displays the dependence on the quartic self-couplings of the LQ doublet (left panel) and LQ singlet (right panel). Row 4 shows the dependence on the quartic couplings between the Higgs doublet and the LQ doublet. Finally, Row 5 shows the dependence on the quartic coupling between the Higgs doublet and the LQ singlet (left panel) and the quartic coupling between the LQ singlet and LQ doublet (right panel).}
    \label{fig:GWpeaksModelParameters}
\end{figure}

In \cref{fig:GWpeaksModelParameters} we present a series of panels illustrating predictions for the SGWB peak amplitude and frequency in terms of various model parameters such as the three LQ physical masses, the trilinear coupling $a_1$, and the various quartic couplings. The Higgs self-coupling, $\lambda_H$, is fixed by the matching procedure and only displays minimal variation across the parameter space; similarly, the Higgs doublet mass parameter $\mu_H$ is fixed by matching to the SM. It is also important to note that the mass terms $\mu_S$ and $\mu_R$ are correlated with their corresponding physical mass parameters.

Our analysis reveals interesting trends in the LQ mass parameters and the trilinear coupling $a_1$. In the region of parameter space that satisfies the high-temperature perturbativity constraint (\cref{eq:highTcheck}), we observe a preference for LQ masses of approximately $\mS\ , \mS2 \approx 2.7~\mathrm{TeV}$ and $\mS1 \approx 2.2~\mathrm{TeV}$.  However, this trend changes in the region associated with the strongest GW peak amplitudes, which corresponds to the region of parameter space that is within the sensitivity range of the planned DECIGO and BBO detectors as well as LISA. In this latter region, the LQ masses tend to take on values around $1.5$ TeV. A similar pattern is observed for the trilinear coupling $a_1$.  In the region of parameter space where the high-temperature perturbativity condition is strictly satisfied, the values of $a_1$ tend to be approximately $1~\mathrm{GeV}$. In contrast, in the region associated with the strongest GW signals, the values of $a_1$ are significantly larger, typically of order $10^3~\mathrm{GeV}$.

Our analysis reveals distinct trends in the quartic couplings.  Stronger phase transitions tend to be associated with larger values of the self-couplings $\lambda_R$ and $\lambda_S$, typically of order $1$ and $10^{-1}$, respectively.  The mixed quartic couplings, however, display a different pattern.  Within the parameter space region where the high-temperature perturbativity condition is strictly satisfied, the mixed quartic couplings typically assume smaller values, generally on the order of $10^{-1}$.  This trend changes as the sensitivity of LISA is approached; in this region, larger values of the mixed quartic couplings, generally of order 1, are favored.  The coupling $g_{HR}$ represents an exception to this trend.  Further analysis of the quartic couplings within the high-temperature perturbative region reveals a distinct preference for the sign of several couplings.  Specifically, the coupling $g_{HR}'$ shows a strong preference for negative values, whereas the couplings $g_{HS}$ and $g_{RS}$ are mostly positive.
\begin{figure}[!ht]
    \centering
    \includegraphics[width=0.6\linewidth]{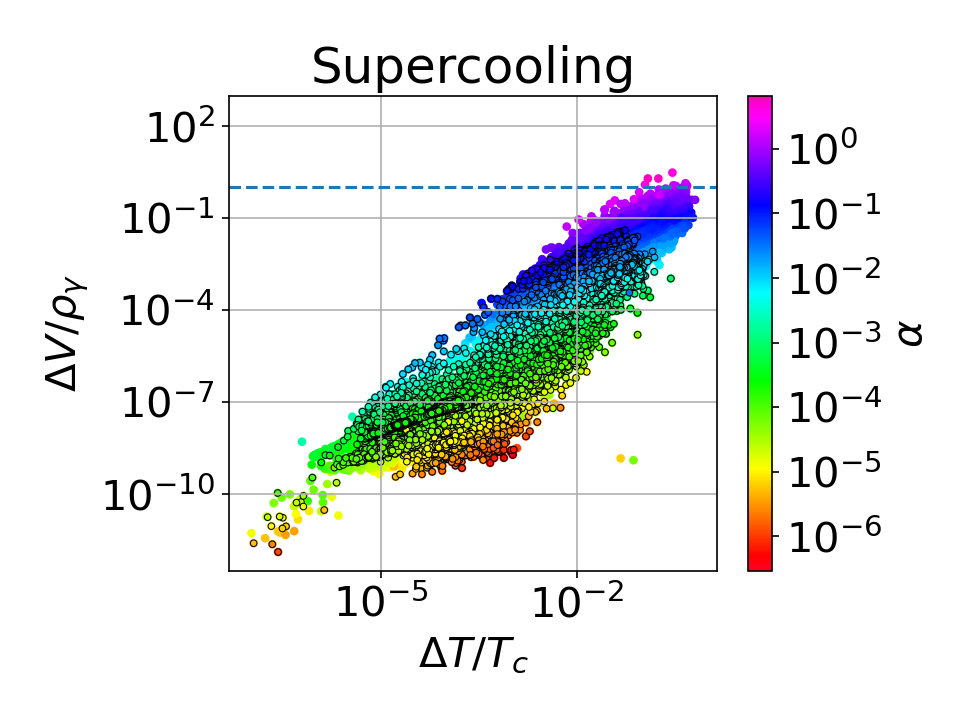}
    \caption{\footnotesize Relation between the supercooling parameter, defined as $(T_c - T_n)/T_c$, and the ratio of vacuum energy density to radiation energy density, $\Delta V/\rho_\gamma$.  A horizontal line is included to identify the boundary between radiation-dominated transitions (below the line) and vacuum-dominated transitions (above the line).  The color scheme of the data points represents the strength of the corresponding phase transitions, as parameterized by $\alpha$.}
    \label{fig:supercooling}
\end{figure}

We conclude this section by discussing the degree of supercooling characterizing the FOPTs identified in our numerical analysis. We quantify the amount of supercooling using the parameter $\Delta T/T_c$, with $\Delta T\equiv T_c - T_n$ the difference between the critical and the nucleation temperatures. The distribution of this supercooling parameter, along with the ratio of vacuum energy density to radiation energy density ($\Delta V/\rho_\gamma$), is shown in \cref{fig:supercooling}.  A striking feature of this distribution is that the vast majority of points correspond to phase transitions exhibiting minimal supercooling, $i.e.$, $\Delta T/T_c \ll 1$. This trend persists even for the strongest transitions identified in our scans, characterized by $\alpha > 1$, where the degree of supercooling remains extremely small ($\Delta T/T_c < 10^{-3}$).  However, it is noteworthy that, while uncommon, the model does allow for the occurrence of phase transitions with substantial supercooling. A dedicated study of such cases is, however, beyond the scope of the present work. It is important to note that the value of $\alpha$ is predominantly determined by the value of $\Delta V/\rho_\gamma$, particularly as one approaches the right edge of the sampled parameter space ($i.e.$, at larger values of $\Delta T/T_c$). However, as one moves towards the left edge of this parameter space ($i.e.$, at lower values of $\Delta T/T_c$), the contribution from entropy injection, which is represented by the second term in equation \cref{eq:alpha}, becomes increasingly relevant in determining the overall value of $\alpha$.

\subsection{Vacuum configurations}\label{sec:vac_config}

This section presents a classification scheme for the various phase transition patterns identified in our analysis. This classification is based on which of the three VEVs, namely $H$, $R$ and $S$ in \cref{eq:vevConfiguration}, acquire non-zero values in each of the phases involved in the transition.  The relationships between these distinct phase transition patterns and the underlying model parameters are then explored.

Regarding the vacuum configurations, two crucial observations are noteworthy:
\begin{outline}[enumerate]
    \1 The extension of the SM through the inclusion of LQs allows, via modifications to the Higgs sector, for \emph{first-order} EWPTs. These transitions can be either from a symmetric state to one where the Higgs field acquires a non-zero VEV ($0 \to h$) or from a state where the Higgs field already has a non-zero VEV to another state with a different Higgs VEV, which we denote ($h \to h$).
    \1 In addition, the model allows for a variety of phase transitions in which the $\SU{3}{C}$ color symmetry is broken in one or both of the phases involved in the transition. 
\end{outline}
To clarify our notation, a phase transition is denoted as $\mathcal{F} \to \mathcal{T}$, where $\mathcal{F}$ refers to the false vacuum configuration and $\mathcal{T}$ refers to the true vacuum configuration.  The labels $h$, $r$, and $s$ are used to indicate the presence of non-zero field values for the corresponding fields ($h$, $r$, $s$) in each phase. A fully symmetric phase, in which all three fields have a value consistent with zero within numerical uncertainties, is indicated by the label $0$.

\begin{figure}[!ht]
    \begin{subfigure}[t]{.60\textwidth}
        \centering
        \includegraphics[width=1.\linewidth]{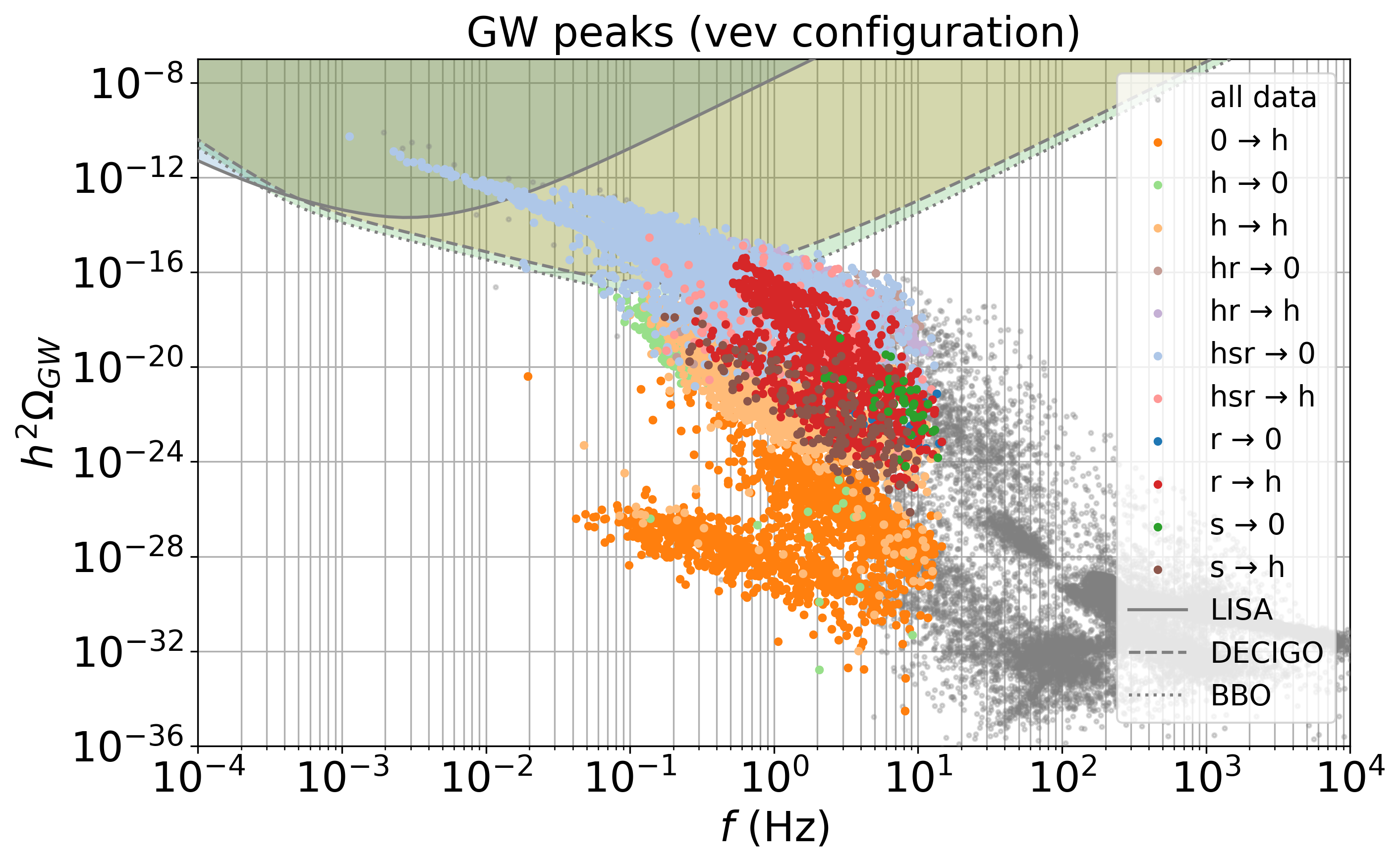}
        \caption{\footnotesize GW peak amplitude and frequency illustrating the various identified vacuum configurations.}
        \label{fig:GWpeaks_vevs}
    \end{subfigure}
    \hfill
    \begin{subfigure}[t]{.35\textwidth}
        \centering
        \includegraphics[width=.9\linewidth]{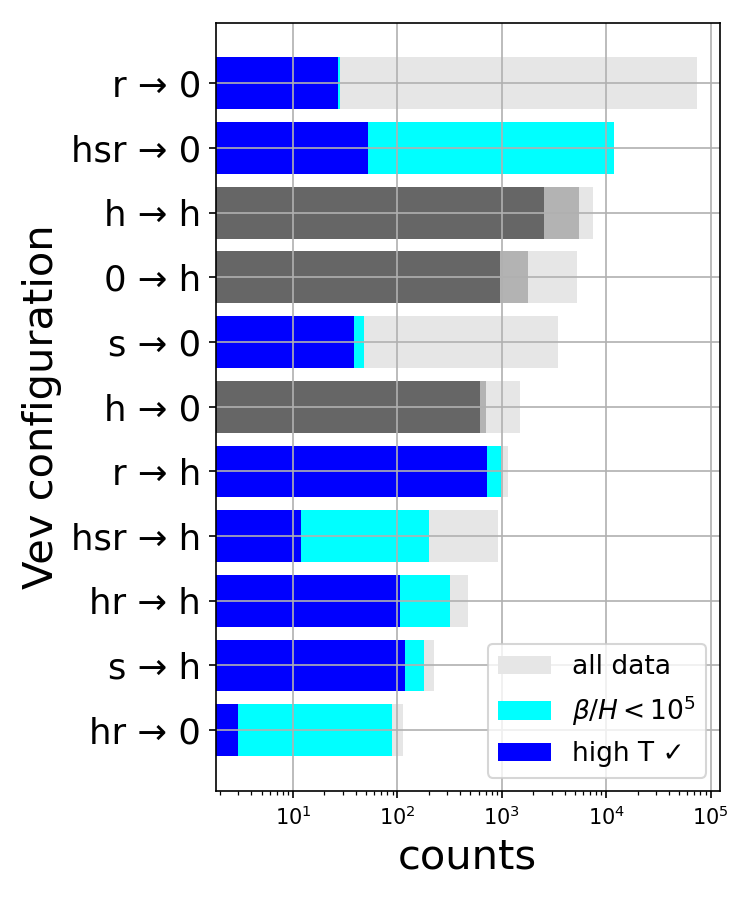}
        \caption{\footnotesize Number counts of the identified \emph{low-CoP} vacuum configurations. $\SU{3}{C}$-preserving transitions are identified by the darker shades of gray.}
        \label{fig:vevDistribution}
    \end{subfigure}
    \caption{\footnotesize Distribution of true vacuum color-preserving (low-CoP) phase transitions. The left panel displays these transitions in the SGWB peak amplitude and peak frequency. Only those transitions for which the condition $\beta/H < 10^5$ is satisfied are shown in color. The high-temperature perturbativity condition is not indicated in this panel for clarity. The right panel shows a histogram indicating the number of occurrences of each of the distinct transition types identified in our analysis.}
    \label{fig:vevs}
\end{figure}

Concerning the first point, although the LQ model does indeed allow for first-order EWPTs of the type $0 \to h$, which are of significant interest in the context of EW baryogenesis \cite{Anderson:1991zb, Sakharov:1967dj, Morrissey:2012db}, these transitions are generally weak in terms of the parameter $\alpha$, typically reaching values of only $\alpha \approx 3 \times 10^{-3}$. This corresponds to a GW peak amplitude of approximately $h^2\Omega_{\rm GW}^{\text{peak}} \approx 10^{-21}$. In marked contrast, all of the strongest phase transitions identified in our analysis, $i.e.$ those that are within the sensitivity range of LISA or future planned gravitational interferometers, are characterized by the spontaneous breaking of the $\SU{3}{C}$ color symmetry in one or both of the phases involved in the transition. The nature of the vacuum configurations for the strong transitions is clearly illustrated in \cref{fig:GWpeaks_vevs}, where the SGWB peak amplitude is color-coded according to the corresponding vacuum configuration.  To the best of our knowledge, this work represents the first numerical analysis of a model that features color-breaking transitions in the early Universe with testable predictions at GW experiments.

\begin{figure}[!ht]
    \begin{subfigure}[t]{.48\textwidth}
        \centering
        \includegraphics[width=1.\linewidth]{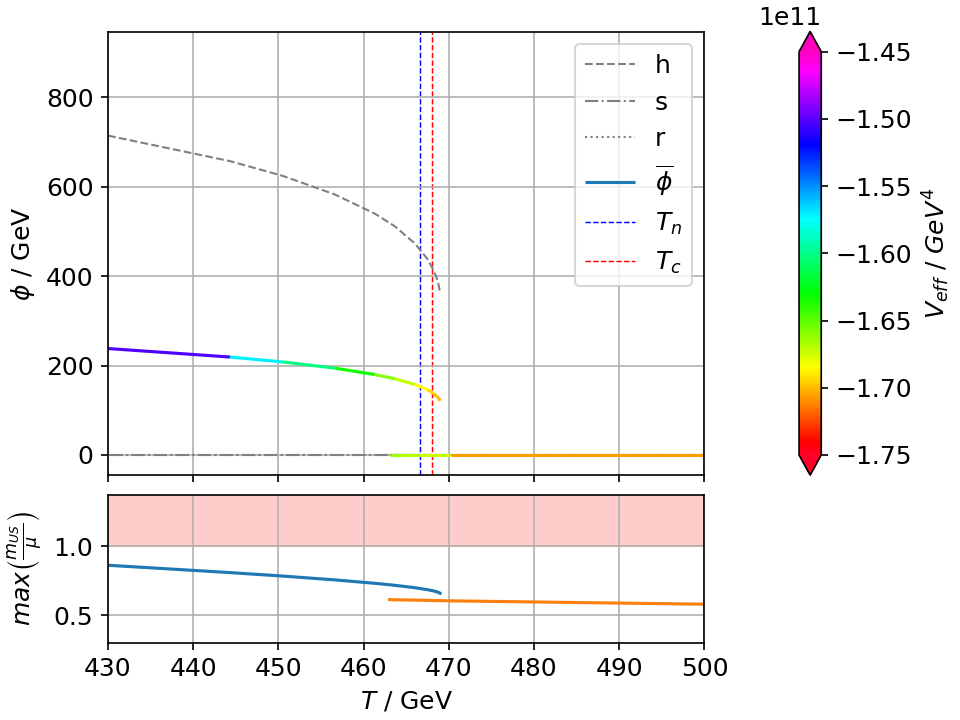}
        \caption{\footnotesize Example of a color-preserving (CoP) phase transition with VEV configuration $0\to h$.}
        \label{fig:phaseDiagram_0-h}
    \end{subfigure}
    \hfill
    \begin{subfigure}[t]{.48\textwidth}
        \centering
        \includegraphics[width=1.\linewidth]{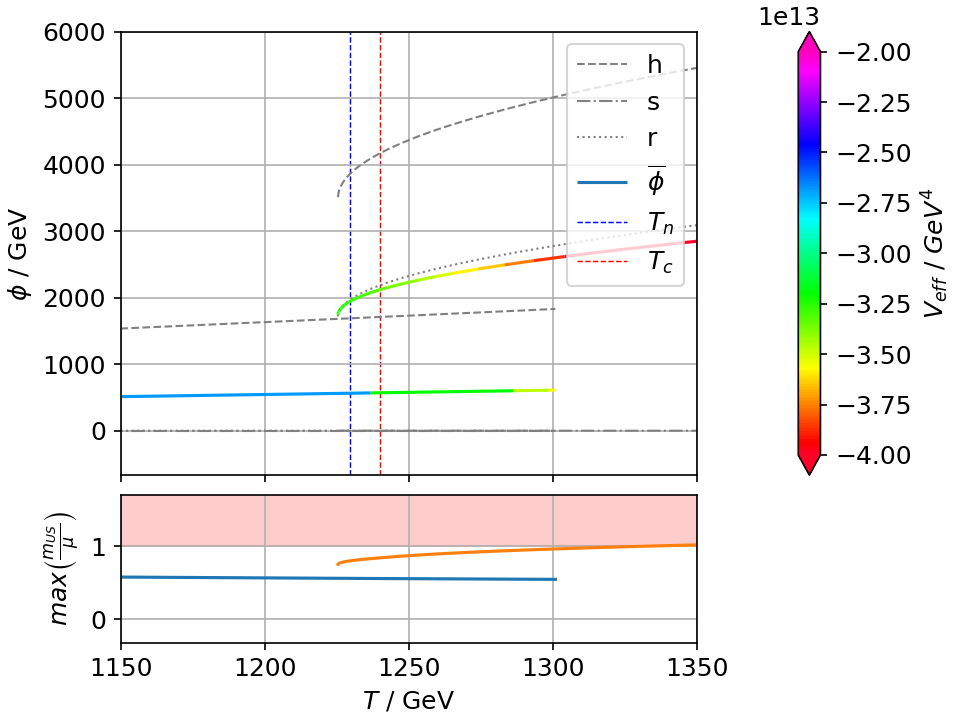}
        \caption{\footnotesize Example of a color-breaking (CoB) phase transition with VEV configuration $(r,h)\to h$.}
        \label{fig:phaseDiagram_rh-h}
    \end{subfigure}
    \caption{\footnotesize Typical phase diagrams of the LQ model. The continuous colored lines represent the mean values of the scalar fields along the path connecting the true and false vacua. The color of these lines is determined by the value of the effective potential at that point along the path.  The thermal VEVs for the three scalar fields, $h$, $s$, and $r$, are shown as gray lines, using different line styles to distinguish between them. The lower panels show the perturbativity condition across the temperature range considered. The red shaded region indicates where perturbativity is not strictly satisfied but remains possible.}
    \label{fig:phaseDiagrams}
\end{figure}

Given the observations above, we adopt a classification scheme for the phase transitions identified in our analysis. This scheme divides the transitions into two main categories: color-preserving (CoP) and color-breaking (CoB). This classification is independent of whether the $\SU{3}{C}$ color symmetry is broken in the high- or low-temperature phase.  Typical examples of phase diagrams for each category are shown in \cref{fig:phaseDiagrams}. These examples include a color-preserving EWPT, denoted as $0 \to h$, and a color-breaking phase transition, denoted as $(r,h) \to 0$. In this latter type of transition, the color symmetry is spontaneously broken in the high-temperature phase due to a non-zero VEV for the field $r$, and is restored in the true vacuum.

To construct the phase diagrams shown in \cref{fig:phaseDiagrams}, we utilize an averaging procedure to represent each phase, which is necessary given the presence of multiple vacuum directions in the model's parameter space. To allow a simultaneous representation of both the temperature and the effective potential, we use the arithmetic average of the temperature-dependent VEV to represent each phase. This averaged VEV is displayed graphically as a continuous curve, the color of which represents the value of the dimensionally reduced effective potential at that temperature\footnote{It is important to note that the potential has units of [mass]$^4$, as per \cref{eq:V3_2_V4}.}.  To facilitate the identification of the VEVs for each of the three scalar fields, $h$, $s$, and $r$, gray lines are also included. The line styles of the gray lines are distinct for each of the three scalar fields. Vertical lines are included in the phase diagrams to indicate the characteristic transition temperatures.

The region of parameter space of primary interest for our analysis spans the range of temperatures from the critical ($T_c$) to the percolation temperature ($T_p$). However, to provide a more comprehensive assessment of the validity of the high-temperature perturbative approach employed in this analysis, we also examine the value of the high-temperature perturbativity parameter defined in \cref{eq:highTcheck} across a wider range of temperatures. This is shown in the panels presented below the phase diagrams. In each of these plots, a curve is shown that represents the maximum value of the ratio of the effective mass to the energy scale for the corresponding phase, displayed as a function of temperature.  It is important to emphasize that, for all of the phase diagrams displayed in \cref{fig:phaseDiagrams}, the high-temperature perturbativity condition is satisfied across the entire considered temperature range. The red shaded area represents the region where the perturbativity condition is not strictly obeyed but is still possible.

Our analysis now focuses on phase transitions that retain color symmetry in the low-temperature phase (low-CoP transitions). In general, it is conceivable that a secondary phase transition at a lower temperature could also restore this symmetry, ensuring consistency with the SM at low energies. However, the high-temperature effective potential employed in this analysis, which is obtained using dimensional reduction \cref{eq:Veffective}, is only valid at sufficiently high temperatures. Extrapolating this effective potential to lower temperatures, such as those near or below the EW scale, is not reliable because we move too far from the matching scale, resulting in large values of the couplings and a possible loss of perturbativity.  Indeed, the effective potential diverges at $T=0$. Therefore, for a conservative and reliable analysis, we restrict our attention to the low-CoP transitions.

Phase transitions that lead to a completely symmetric vacuum, characterized by zero VEVs for all three scalar fields ($0,0,0$), represent a reasonable scenario.  This is because the SM predicts a crossover phase transition from a symmetric phase to a phase with a non-zero Higgs VEV ($0 \to h$) in the vicinity of the EW scale.  Our LQ model is explicitly constructed to match the SM in this regime; the specific matching conditions are given in \cref{eq:matching_mu,eq:matching_lambda}.  To verify this matching, we have explicitly minimized the tree-level scalar potential given in \cref{eq:Vtree}. We confirm that this minimization procedure consistently yields a value for the VEV that agrees with that in the SM. 

Figure \cref{fig:GWpeaks_vevs} displays the distribution of low-CoP vacuum configurations in the plane of GW peak amplitude versus peak frequency.
Examination of this distribution reveals that the strongest SGWB signal predictions are associated with phase transitions of the type $(h,s,r) \to 0$, followed by transitions of the type $(h,s,r) \to h$ and $r \to h$. Among these, the latter feature the largest number of points that strictly satisfy the high-temperature perturbativity condition described in section \ref{sec:high-T}. The $(h,s,r) \to 0$ transitions comprise the majority of points satisfying the constraint $\beta/H < 10^5$.  However, it is important to note that the strongest transitions among those of the $(h,s,r) \to 0$ type exhibit values of the perturbativity parameter $m_{\text{US}}/\mu \approx 2$. A fraction of these points fall within the sensitivity range of LISA, suggesting the potential for constraining a portion of the model's parameter space using next-generation GW detectors.

The right-hand panel of \cref{fig:vevs} presents a histogram showing the distribution of the various low-CoP vacuum configurations obtained in our analysis.  For clarity, the histogram distinguishes between those transitions for which the condition $\beta/H < 10^5$ is satisfied, and those where the high-temperature perturbativity condition given in \cref{eq:highTcheck} is also satisfied.  The latter is discussed in more detail in sections \ref{sec:GWpeaks} and \ref{sec:high-T} of this work. Darker shades of gray represent vacuum configurations that preserve color symmetry throughout the transition.

Based on the considerations outlined above, our analysis has identified a total of 11 distinct types of phase transitions that are physically viable ($i.e.$ color is preserved in the low phase). These 11 transition types are selected from a total of 64 possible configurations. Among these 11 viable types, three correspond to transitions that are purely EW in nature and do not involve the breaking of the color symmetry. The remaining eight viable types involve phase transitions in which the color symmetry is spontaneously broken in the high-temperature phase.

\subsection{Improving the accuracy of SGWB predictions}

Theoretical uncertainties are inherent to several steps of our thermodynamic analysis. The estimation of the energy budget available to the sound waves produced during bubble collisions relies on simplified models that assume a bag equation of state \cite{Ellis:2018mja}. This simplification impacts other parameters, such as the bubble wall velocity ($\xi_w$). Further approximations are made in the calculation of the decay rate.  Specifically, the prefactor in the expression for the decay rate given in \cref{eq:decayRate} ($T^4 e^{-S_3/T}$) represents a simplified approximation of the model-specific functional determinants that should appear in the exact expression for the decay rate. While a numerical tool for calculating these functional determinants is available \cite{Ekstedt:2023sqc}, its current implementation is limited to the case of a single scalar field. The development of a multi-field version of this tool would permit a more precise calculation of the prefactor in the decay rate and would constitute a natural and important extension of the current work.  Until such a tool is available, however, we must rely on the approximate expression for the decay rate given in \cref{eq:decayRate}.

The thermodynamic properties of the LQ model were analyzed using the dimensional reduction approach. This approach offers significant advantages in terms of reliability at high temperatures compared to alternative methods \cite{Braaten:1995cm,Kajantie:1995dw,Croon:2020cgk,Lewicki:2024xan}. The effective potential for the model was calculated at NLO using the \texttt{DRalgo} software package.  While \texttt{DRalgo} is capable of performing calculations at NNLO, we have focused on an NLO treatment. This choice was primarily driven by computational considerations.  Specifically, the numerical calculation of thermodynamic quantities at NNLO requires tracing the phases and solving the bounce equation in a three-field space ($h$, $\red r$, $\blue s$), which is computationally very expensive and will be done elsewhere. 

\section{Summary and conclusions} \label{sec:conclusions}

This article presents a detailed analysis of the primordial GW spectrum arising from FOPTs in the minimal LQ model defined by the scalar potential given in \cref{eq:Vtree}. This particular model has been shown in a previous work \cite{Freitas:2022gqs} to be consistent with existing experimental constraints on flavor physics and also capable of providing a mechanism for the generation of Majorana masses for active neutrinos. In this work, we explore the phase transition dynamics in this model. Our analysis reveals that multiple viable FOPT patterns exist, characterized by different configurations of the true and the false vacua.

The minimal LQ model under consideration in this study exhibits a unique combination of the following features:
\begin{itemize}
    \item First, it allows for the occurrence of first-order EWPTs, in which only the Higgs field acquires a non-zero VEV.
    \item Second, the model also permits phase transitions in which the $\SU{3}{C}$ color symmetry is broken spontaneously, either in the high-temperature phase, the low-temperature phase, or both.
    \item The detectability region is characterized by physical masses of approximately $1.5~\mathrm{TeV}$ for all three LQs, consistent with the current collider bounds and within the reach of the LHC (both at run-3 and the high-luminosity upgrade) for further investigation.
    \item The observability region places strong constraints on the the scalar potential.  Specifically, it predicts values for the various quartic couplings of order $10^{-2}$ with the exception of $\lambda_R$ which takes values of order $1$.
    \item The trilinear coupling $a_1$, relating all three scalar fields $H,R$ and $S$, plays a crucial role throughout this work. Besides providing mixing between the $S$ singlet and one of the components of the $R$ doublet, thus allowing for the radiative generation of neutrino masses and their mixing structure, it tends to assume values of approximately $2~\mathrm{TeV}$ in the detectability region, while acquiring values of $\order{1~\mathrm{GeV}}$ in the adjacent regions.
\end{itemize}

Our study indicates that a significant portion of the model's parameter space may be within reach of future GW detectors, particularly LISA, if we allow for the perturbativity condition $m_{\text{US}}/\mu$ to be slightly above unity. Specifically, we find that values around $m_{\text{US}}/\mu \approx 2$ are sufficient. Note that the criterion for high-temperature perturbativity given in \cref{eq:highTcheck} should not be considered a strict threshold.  Determining the precise boundary of the perturbative regime for a given model is a non-trivial task beyond the scope of this first analysis.

Despite the potential for future refinements in the accuracy of primordial GW spectrum calculations, this work offers a substantial improvement in our understanding as demonstrated in \cref{fig:GWspectra_scaleFactor}. This work is, to the best of our knowledge, the first numerical study to explore a model that simultaneously incorporates a mechanism for the generation of neutrino masses and the phenomenon of color-restoration FOPTs in the early Universe, producing an observable SGWB detectable by future planned GW facilities such as LISA, BBO, and DECIGO. Our analysis provides concrete predictions for various model parameters, testable at colliders, specifically within the region that leads to detectable GW signals.

\cleardoublepage

\section*{Acknowledgements}
We thank Andreas Ekstedt for several insightful discussions about dimensional reduction. We thank Jo\~ao Gon\c{c}alves for useful discussions on the leptoquark model and the BSM-to-gravitational wave pipeline throughout the project, and Vasileios Vatellis for his assistance during the initial stages. M.F.\ and A.P.M.\ are supported by the Center for Research and Development in Mathematics and Applications (CIDMA) through the Portuguese Foundation for Science and Technology (FCT - Funda\c{c}\~{a}o para a Ci\^{e}ncia e a Tecnologia), references UIDB/04106/2020 (\url{https://doi.org/10.54499/UIDB/04106/2020}) and UIDP/04106/2020 (\url{https://doi.org/10.54499/UIDP/04106/2020}). A.P.M.\ and  M.F.\ are also supported by the projects with references CERN/FIS-PAR/0019/2021 (\url{https://doi.org/10.54499/CERN/FIS-PAR/0019/2021}), CERN/FIS-PAR/0021 /2021 (\url{https://doi.org/10.54499/CERN/FIS-PAR/0021/2021}) and CERN/FIS-PAR/0025/2021 (\url{https://doi.org/10.54499/CERN/FIS-PAR/0025/2021}).
M.F.\ is also directly funded by FCT through the doctoral program grant with the reference PRT/BD/154730/2023 within the scope of the ECIU University.
A.P.M.~is also supported by national funds (OE), through FCT, I.P., in the scope of the framework contract foreseen in the numbers 4, 5 and 6 of the article 23, of the Decree-Law 57/2016, of August 29, changed by Law 57/2017, of July 19 (\url{https://doi.org/10.54499/DL57/2016/CP1482/CT0016}).
R.P., M.B. and J.R.~are supported in part by the Swedish Research Council grant, contract number 2016-05996, as well as by the European Research Council (ERC) under the European Union's Horizon 2020 research and innovation programme (grant agreement No 668679). R.P.~also acknowledges support by the COST Action CA22130 (COMETA).

\appendix

\section{Dimensional reduction: effective theories at high temperature}\label{Ap:DRalgo}

Perturbation theory converges slowly at finite temperatures. The problem at hand features two energy scales: the temperature and the scalar mass. Since these scales are vastly different, it is apt to use an effective theory by integrating out high-energy modes with $E \sim  T$. Since the system is in equilibrium, there is no time dependence, leaving an effective (spatial) three dimensional field theory. The procedure to obtain such a theory, outlined below, is referred to as \emph{dimensional reduction} (DR).

In the imaginary time formalism we let $t\rightarrow -i\tau$, where $\tau\in [-\beta,+\beta]$, with $\beta=1/T$. Because the bosonic and fermionic fields are periodic and anti-periodic, respectively, in $\tau$, the fields and their second derivatives can be Fourier expanded according to
\begin{align}
    &\phi(\tau,\Vec{x})=T\sum_{n=-\infty}^\infty e^{i \tau \omega_n} \phi_n(\Vec{x})
    \\& \partial^2 \phi(x) = T \sum_{n=-\infty}^\infty e^{i \tau \omega_n}\left[\omega_n^2\phi_n(\Vec{x})-\Vec{\nabla}^2\phi_n(\Vec{x})\right]\,,
\end{align}
where $\omega_n$ are the so-called Matsubara frequencies, with $\omega_n = 2n \pi T$ for bosons and $\omega_n = (2n+1) \pi T$ for fermions. Consequently, we see that there is an infinite tower of modes with squared masses $m_n^2 = \omega_n^2+m^2$, with $m$ denoting the ordinary mass. When $T$ is large, in the sense that $T \gg m$, all the fermionic modes as well as the non-zero bosonic modes are heavy, and can be integrated out by matching correlators.

The couplings of the effective Lagrangian are matched to those in the 4d theory, and depend implicitly on the temperature. To avoid large logarithms, these couplings are matched at a scale $\mu \sim T$:
\begin{align}
    \lambda_\text{3d}=T\left(\lambda_\text{4d}+a g_\text{4d}^4 \log\frac{\mu}{T}+b g_\text{4d}^4 +\ldots \right)\,.
\end{align}
Since the couplings are independent of the matching scale --- $\mu \partial_\mu \lambda_\text{3d}=0$ --- $\mu\sim T$ is chosen to minimize the logarithm. In practice, this involves starting with a theory at a given scale $\mu_0$, evolving the couplings to $\mu \sim T$, and calculating observables with the effective theory.

In the 3d EFT thus constructed, said to be living at the \emph{soft} scale, the temporal components of the vector fields appear as decoupled scalar fields, with associated masses called Debye masses. In practice, and this is an assumption we make throughout the present project, these Debye masses are often larger than the scale of the phase transition. Hence, we are justified in further integrating out the temporal modes, to construct yet another 3d EFT, now said to be living at the \emph{ultrasoft} scale. The \texttt{DRalgo} \cite{Ekstedt:2022bff} package constructs the soft and ultrasoft 3d EFT in two steps:
\begin{outline}
    \1 \textbf{Dimensional reduction:}\\
    The 4d fundamental theory, living at the \emph{hard} energy scale $\mu_{4d} \sim T$, is matched to a 3d \emph{soft} theory, with renormalization scale $\mu_{\text{3d}}^{\text{S}} \sim gT$. This removes logarithms of the form $\log{\frac{\mu}{T}}$ for energy scales $\mu \sim T$. At this stage, 4d $\beta$-functions, thermal masses, effective couplings and anomalous dimensions are computed.
    \1 \textbf{Integrating out massive temporal scalars:}\\
    The soft theory is matched to an \emph{ultrasoft} theory at the energy scale $\mu_{\text{3d}}^{\text{US}} \sim g^2 T$, removing logarithms of the form $\log{\frac{m_D}{\mu}}$, with $m_D$ denoting a typical Debye mass.
    Here, 3d $\beta$-functions, masses and couplings are computed. This step is optional and user-defined\footnote{
    See \cite{Gould:2023ovu,Kierkla:2023von,Lewicki:2024xan} for studies addressing the soft and intermediate scales.
    }.
\end{outline}
The matching between the 3d EFT and the underlying 4d theory is performed to next-to-leading order (NLO). At this level, two-loop thermal corrections to the masses of the scalar fields are included. All couplings are resummed to one-loop order. The \texttt{DRalgo} package also allows for integrating out additional heavy scalars. For a comprehensive description of \texttt{DRalgo}'s capabilities and implementation, we refer to \cite{Ekstedt:2022bff}.

The high-temperature theory can also be used to calculate the thermal escape rate. This is determined by the probability for the scalar fields to overcome the potential barrier separating the metastable vacuum from the true vacuum and is set by the Boltzmann factor $e^{-S_3/T}$. To determine the bounce action it is sufficient to expand the path integral around a saddle point rather than performing a full dynamical treatment. This bounce solution satisfies \cref{eq:bounce}, with the boundary conditions \cref{eq:bounceBoundCond}.
The rate for thermal escape is then given by \cref{eq:decayRate}.

As mentioned in \cref{sec:numerics}, for the implementation we use \texttt{Dratopi} \cite{Dratopi}, a soon-to-be-released tool which interfaces the \texttt{DRalgo} package with \texttt{Python} and a modified version of \texttt{CosmoTransitions} \cite{Wainwright_2012}. \texttt{Dratopi} provides a script to export from \texttt{DRalgo} into \texttt{Python}, among other things, the beta functions for the 4d theory, the results of the hard-to-soft (step 1 above) and the soft-to-ultrasoft (step 2 above) matchings, as well as the effective potential in the ultrasoft 3d EFT. Moreover, \texttt{Dratopi} also provides the necessary routines to calculate the 4d thermal effective potential and to use this for phase transition analysis in a slightly modified version of \texttt{CosmoTransitions}. Below, we summarize the main steps to obtain the 4d thermal effective potential $V_\text{eff}^\text{4d}$ at field values $v^\text{4d}$ and temperature $T$.
Note that the first two steps are executed only once, upon initialization of the model.

\begin{enumerate}
    \item Define the model by specifying its 4d parameters, collectively denoted $\textbf{p}_{\text{4d}}$, at some given reference energy scale $\mu_0$.
    \item Using the beta functions, solve the renormalization group (RG) equations, to obtain an interpolated solution of $\textbf{p}_{\text{4d}}(\mu)$ as a function of the RG scale/energy scale $\mu$ (over some specified range).
    \item Set the hard matching scale to $\mu_{4d} = C_{\text{4d}} \cdot T$, where $C_{\text{4d}}$ is a prefactor that defaults to $\pi$. \footnote{In this project, we normally use the default value of $C_{\text{4d}} = \pi$. As discussed above, in connection with \cref{fig:GWspectra_scaleFactor}, we also consider the influence of changing $C_{\text{4d}}=\mu_{\text{4d}}/T$.}
    \item Construct the soft 3d EFT, by matching the 4d theory to the 3d EFT, at the scale $\mu_{\text{4d}}$.
    \item Set the soft-to-ultrasoft matching scale to $\mu_{\text{3d}}^{\text{US}}=C_{\text{3d}} \cdot m_{\text{D,min}}$. Here, $m_{\text{D,min}}$ is equal to the smallest Debye mass, at temperature $T$. Further, $C_{\text{3d}}$ is a prefactor which defaults to 1. \footnote{In this project, we do not consider the impact of varying $C_{\text{3d}}$.}
    \item Construct the ultrasoft 3d EFT, by integrating out the temporal modes, thus obtaining the 3d parameters $\textbf{p}_{\text{3d}}^\text{US}(T)$ in the ultrasoft 3d EFT, at the given temperature.
    \item Calculate the 4d thermal effective potential according to the relation $V_{\text{eff}}^{\text{4d}}(v^\text{4d},T) = TV_\mathrm{eff}^{\text{3d}}(v^\text{4d}/\sqrt{T};\textbf{p}_{\text{3d}}^\text{US}(T))$. Here, $V_\mathrm{eff}^{\text{3d}}$ is the effective potential in the ultrasoft 3d EFT, calculated at field values $v^\text{3d}=v^\text{4d}/\sqrt{T}$ and at 3d ultrasoft parameter values $\textbf{p}_{\text{3d}}^\text{US}(T)$.
\end{enumerate}

Except for the input parameters (step 1.), these steps are fully automatized by the package. In summary, \texttt{Dratopi} enables users to leverage the advantages of dimensional reduction with minimal intervention, providing a streamlined and efficient framework for studying cosmological FOPTs.

\section{Monte Carlo scanning}

Given the limited number of points satisfying all perturbativity conditions, we perform a simple Monte-Carlo (MC) scan of the parameter space of the theory, initialized on benchmark points from each vacuum configuration. Our MC algorithm consists of three parts:
\begin{outline}[enumerate]
    \1 An initial point is randomly selected from a dataset of ``optimal'' points featuring FOPTs and GW production.
    \1 A \emph{small} random move in the space of quartic parameters as well as all free parameters of the theory, $i.e.$ LQ masses and trilinear couplings $a_1$. This defines the path of a random walker exploring the parameter space of the theory.
    \1 Phase transition (if any) and GW parameters are computed for the new model\footnote{This is carried on with the standard prescription: phase tracing, search for phase transition, computation of transition and GW parameters (if any of these is first order).}.
    An ``improvement'' function evaluates the overall change in quartic parameters and $m_{US}/\mu$ ratio,
    \begin{equation}
        \mathfrak{i} \equiv -\left(a\expval{\Delta \frac{m_{\mathrm{US}}}{\mu}} 
        + b\expval{\Delta \abs{\lambda}}
        \right)\, ,
    \end{equation}
    where $\lambda$ represents all quartic couplings and $\ev{}$ denotes the parameter average ($e.g.$, the average of $m_{\mathrm{US}}/\mu$ at the high and low VEVs), which we only include if at least one parameter exceeds a certain threshold. Furthermore, $a$ and $b$ are arbitrary parameters chosen to optimize the scanning; we set both to $10$. The move is then \emph{accepted} or \emph{rejected} according to the value of the improvement:
    \begin{equation}
        e^\mathfrak{i}\ \left\{\mqty{\ge r & \Rightarrow & \texttt{accept} \\
            < r & \Rightarrow & \texttt{reject}
            } \right.\ ,
    \end{equation}
    where $r$ is a random variable with uniform distribution $U(0,1)$. As a result, moves with positive improvement are always accepted, while those with negative improvement are only accepted with exponentially decreasing probability. We then iterate from bullet point 2, starting from either the newly accepted point or the previous ones.
\end{outline}
The typical path of the MC random walker, in the $\{m_{\text{US}}/\mu, g_{\text{HS}}\}$ space, is shown in \cref{fig:mcRun} below. Statistically, it moves towards optimal values of both parameters.
\begin{figure}[!htb]
    \centering
    \includegraphics[width=0.75\linewidth]{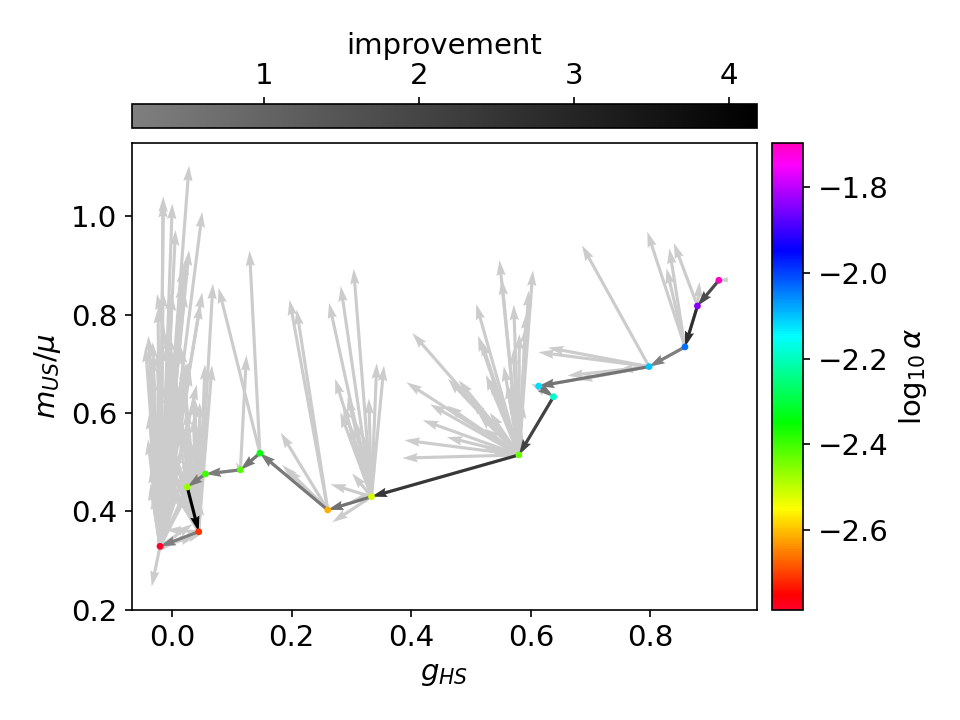}
    \caption{\footnotesize A typical path in the $\{m_{US}/\mu, g_{HS}\}$ space of the random walker implemented in our Monte Carlo algorithm. Points are color-coded according to the value of the $\Omega_{\rm GW}$ peak (right bar). Grey to black arrows (those landing on a colored point) identify the MC accepted steps -- with color representing the overall ``improvement'' (top bar) -- while light grey arrows represent rejected steps.}
    \label{fig:mcRun}
\end{figure}

\newpage
\bibliography{bib}

\end{document}